\documentclass[a4paper,fleqn,usenatbib]{mnras}

\usepackage{mathptmx}

\usepackage[T1]{fontenc}
\usepackage{ae,aecompl}

\usepackage{graphicx}	
\usepackage{amsmath}	
\usepackage{amssymb}	
\usepackage{natbib}

\newcommand{\cahk}{$\rm CaH\&K$}
\newcommand{\mgf}{$\rm Mg4780$}
\newcommand{\nad}{$\rm NaD$}
\newcommand{\tioi}{$\rm TiO1$}
\newcommand{\tioiio}{$\rm TiO2_{SDSS}$}

\newcommand{\cai}{$\rm Ca1$}
\newcommand{\caii}{$\rm Ca2$}

\newcommand{\caiii}{$\rm Ca3$}
\newcommand{\hbo}{$\rm H\beta_o$}
\newcommand{\hb}{$\rm H\beta$}
\newcommand{\mgb}{$\rm Mgb5177$}
\newcommand{\mgfep}{$\rm [MgFe]'$}
\newcommand{\mgfe}{$\rm [Mg/Fe]$}
\newcommand{\gammab}{$\rm \Gamma_b$}

\title[Galaxy Environment vs IMF of ETGs]{The Influence of Galaxy Environment on the Stellar Initial Mass Function of Early-Type Galaxies}

\author[Rosani et al.]{
Giulio Rosani,$^{1,2}$\thanks{E-mail: rosani@astro.rug.nl}
Anna Pasquali,$^{1}$
Francesco La Barbera,$^{3}$
Ignacio Ferreras,$^{4}$
\newauthor and Alexandre Vazdekis$^{5,6}$
\\
$^{1}$Astronomisches Rechen-Istitut, Zentrum f\"ur Astronomie, Universit\"at Heidelberg, M\"onchhofstr. 12-14, D-69120 Heidelberg, Germany\\
$^{2}$Kapteyn Astronomical Institute, University of Groningen, P.O. Box 800, 9700AV Groningen, The Netherlands\\
$^{3}$INAF - Osservatorio Astronomico di Capodimonte, Salita Moiariello 16, I-80020 Napoli, Italy\\
$^{4}$Mullard Space Science Laboratory, University College London, Dorking, Surrey RH5 6NT, UK\\
$^{5}$Instituto de Astrof\'isica de Canarias, E-38200 La Laguna, Tenerife, Spain\\
$^{6}$Departamento de Astrof\'isica, Universidad de La Laguna, E-38205 La Laguna, Tenerife, Spain
}

\date{Accepted XXX. Received YYY; in original form ZZZ}

\pubyear{2017}

\begin{document}
\label{firstpage}
\pagerange{\pageref{firstpage}--\pageref{lastpage}}
\maketitle

\begin{abstract}
In this paper  we investigate whether the stellar initial mass function of early-type galaxies depends on their host environment. To this purpose, we have selected a sample of early-type  galaxies from the SPIDER catalogue, characterized their environment through the group catalogue of Wang et al. and used their optical SDSS spectra to constrain the IMF slope, through the analysis of IMF-sensitive spectral indices. To reach a high enough signal-to-noise ratio, we have stacked spectra in velocity dispersion ($\sigma_0$) bins, on top of separating the sample by galaxy hierarchy  and host halo mass, as proxies for galaxy environment. In order to constrain the IMF, we have compared observed line strengths to predictions of MIUSCAT/EMILES synthetic stellar population models, with varying age, metallicity, and ``bimodal'' (low-mass tapered) IMF slope ($\rm \Gamma_b$). Consistent with previous studies, we find that $\rm \Gamma_b$ increases with $\sigma_0$, becoming bottom-heavy (i.e. an excess of low-mass stars with respect to the Milky-Way-like IMF) at high $\sigma_0$. We find that this result is robust against the set of isochrones used in the stellar population models, as well as the way the effect of elemental abundance ratios is taken into account. We thus conclude that it is possible to use currently state-of-the-art stellar population models and intermediate resolution spectra to consistently probe IMF variations. For the first time, {\it we show that there is no dependence of $\Gamma_b$ on environment or galaxy hierarchy, as measured within the $3''$ SDSS fibre, thus leaving the IMF as an intrinsic galaxy property, possibly set already at high redshift}.
\end{abstract}

\begin{keywords}
galaxies:elliptical and lenticular,cD -- galaxies:fundamental properties -- galaxies:stellar content -- galaxies:groups:general -- galaxies:evolution
\end{keywords}



\section{Introduction}
\label{sec:intro}

The study of the formation and evolution of early-type galaxies (ETGs) has been carried out for a long time, yet today it still poses some interesting challenges. Today's increasingly accepted scenario for the formation of ETGs is the two-phase scenario \citep{2010gfe..book.....M,2010ApJ...725.2312O,2013IAUS..295..340N}, in which roughly half the final mass of the galaxy is formed in a relatively short starburst phase at high redshift (the formation phase), followed by a second phase, where the other half is accreted over time through galaxy-galaxy interactions such as minor and major mergers (the assembly phase, see also \citealt{2006MNRAS.366..499D}). The properties of the stellar population, formed during the initial starburst, are found to correlate with the central velocity dispersion, hence with the mass, of the galaxy (\citealt{1973ApJ...179..731F,1992ApJ...398...69W,2000AJ....120..165T,2006MNRAS.370.1106G,2009ApJ...693..486G,2009ApJ...698.1590G,2010MNRAS.408...97K,2015MNRAS.448.3484M}, but see also \citealt{2006ARA&A..44..141R} and references therein), with more massive galaxies having higher $\rm [\alpha/Fe]$, indicative of shorter and more intense starbursts, as well as older and more metal-rich stellar populations~\citep{1996ApJS..106..307V,1997ApJS..111..203V,2005ApJ...621..673T}. Radial gradients of age, metallicity and elemental abundances obtained by \citet{2015ApJ...807...11G} indicate that, while these populations dominate the ETGs central regions, the galaxy outskirts are made up of metal-poorer stars. Such metal-poor populations may have been accreted over time from smaller systems \citep{2012MNRAS.426.2300L,2016ApJ...821..114H}. Extended stellar features observed in many ETGs (see \citealt{2015MNRAS.446..120D}) suggest in fact that galaxy interactions are common and simulations show that in the event of a minor merger the stellar content of the less massive galaxy undergoing the merger is deposited in the outskirts of the more massive stellar system. Major mergers on the other hand are capable of mixing the stellar content of both galaxies, but happen in general only once in the lifetime of an ETG \citep{2009MNRAS.395.1491B,2011MNRAS.415.3903T,2012ApJ...744...63O,2013IAUS..295..340N}.\par
An  important factor, regulating the type and rate of mergers that galaxies may have undergone, is the environment where they reside. Following the approach used in semi-analytic models of galaxy formation and evolution (SAMs), environment can be characterized by the mass of the dark matter (DM) host halo that galaxies are bound to. Moreover, these galaxies can be split between the host central galaxy, which is the most massive one, and satellites. Theoretical predictions as well as observations show the evolutionary paths of these two galaxy types to be rather different. Centrals are situated in a spot where the host halo enables them to accrete gaseous and stellar material from satellites, while satellites are being stripped of their stars and gas by tidal and ram-pressure stripping, respectively \citep{1972ApJ...176....1G,2009MNRAS.399.2221B,2009A&A...499...87K,2012MNRAS.424.2401V,2013MNRAS.431.3533C}. In this way the star formation (SF) of centrals is more extended in time, while the SF in satellites is quenched by environment, thus making galaxy hierarchy influence the overall stellar population  properties of galaxies \citep{2009MNRAS.394...38P,2010MNRAS.407..937P,2010MNRAS.405..329R,2011MNRAS.418L..74D,2014MNRAS.445.1977L,2015AN....336..505P}.\par
Representing environment with the halo dark matter mass allows us to correlate galaxy properties with a global measurement of environment and to directly compare observational trends with what is predicted by SAMs. On the contrary, the projected number density of satellites, which is often used in the literature to quantify environment, does not allow such a direct comparison. For example, the projected number density of satellites in a small galaxy group most likely describes the whole environemnt, while it delivers only a measurement of the local environment in the case of galaxy clusters \citep{2009MNRAS.394...38P,2010MNRAS.407..937P,2015AN....336..505P}.\par
Since galaxy environment has been shown to influence the stellar population content of galaxies, a fundamental question is to assess to what an extent different stellar population properties depend on hierarchy and the environment where galaxies reside. In the present work, we focus on one of these properties, i.e. the stellar initial mass function (IMF) of ETGs.\par
In recent years, the stellar IMF of ETGs has been found to deviate significantly from the Galactic function, i.e. either a \citet{2001MNRAS.322..231K} or a \citet{2005ASSL..327...41C} distribution, with growing evidence for an excess of low-mass stars, i.e. a bottom-heavy IMF, in more, relative to less, massive galaxies \citep{1996ApJS..106..307V,1997ApJS..111..203V,2003MNRAS.340.1317V,2003MNRAS.339L..12C,2010Natur.468..940V,2011ApJ...735L..13V,2012ApJ...760...70V,2012ApJ...747...69C,2012ApJ...760...71C,2013MNRAS.429L..15F,2015hsa8.conf..102F,2015MNRAS.448L..82F,2013MNRAS.433.3017L,2015MNRAS.449L.137L,2014MNRAS.438.1483S,2015ApJ...806L..31M,2017ApJ...837..166C}. Such evidence for a non-universal IMF has been confirmed with different observational methods:
\begin{description}
\item[ Dynamics --] The total, dynamical mass (or mass-to-light ratio, M/L) of an ETG is derived, and then the stellar mass (or M$_\star$/L) is inferred, based on some assumption on the underlying dark matter distribution, and compared to the expected value for a Kroupa-like IMF \citep{2011MNRAS.415..545T,2012Natur.484..485C,2013MNRAS.432.1862C,2012MNRAS.422L..33D,2013MNRAS.432.2496D,2012AJ....144...78W,2013ApJ...765....8T,2014ApJ...792L..37M,2017MNRAS.464..453D}.
\item[Spectral analysis -- ] IMF-sensitive  features in the spectra of ETGs are compared to predictions of synthetic stellar population models with varying IMF, either through the analysis of line-strenghts or spectral fitting, to constrain directly the fraction of low-mass stars in the IMF \citep{1962ApJ...135..715S,1978QJRAS..19..177C,1980ApJ...235..405F,1986ApJ...311..637C,1988ApJ...325L..29H,1992AJ....103..711D,2002ApJ...579L..13S,2003ApJ...588L..17F,2003MNRAS.339L..12C,2010Natur.468..940V,2011ApJ...735L..13V,2012ApJ...760...70V,2012ApJ...747...69C,2012ApJ...760...71C,2012ApJ...753L..32S,2014MNRAS.438.1483S,2012MNRAS.426.2994S,2013MNRAS.429L..15F,2015hsa8.conf..102F,2015MNRAS.448L..82F,2013MNRAS.433.3017L,2015MNRAS.449L.137L,2015ApJ...806L..31M,2017ApJ...841...68V}.
\item[Lensing -- ] The total  mass projected within the Einstein radius is measured. Based on assumptions on the dark matter component, the stellar mass is inferred, and compared to expectations (based on photometry/spectroscopy) for a Kroupa-like IMF~\citep{2005ApJ...623L...5F,2008MNRAS.383..857F,2010MNRAS.409L..30F,2010ApJ...724..511A,2010ApJ...709.1195T,2011MNRAS.415.2215B,2015MNRAS.452.2434S,2016arXiv161200065N}. This method differs from dynamics not only in the techniques used to constrain the total mass (or M/L), but also in that it constrains the 2D projection of the mass of the galaxy on the lens plane and not the 3D distribution of the mass as dynamical studies do.
\end{description}
While, in principle, spectroscopy allows the IMF shape to be directly constrained \citep{2012ApJ...747...69C}, lensing and dynamics do actually constrain only the IMF normalization (i.e. the stellar mass), which is affected by either low-mass stars (i.e. a bottom-heavy distribution) or stellar remnants (i.e. a top-heavy distribution, with an excess of giant, relative to dwarf, stars relative to the Milky-Way distribution). Moreover, some works have found evidence for a Kroupa-like IMF normalization in some massive ETGs, leaving the debate on the IMF slope in ETGs open (see \citealt{2013MNRAS.434.1964S,2015MNRAS.449.3441S,2016MNRAS.459.3677L}).\par
From a theoretical point of view, there is no commonly accepted framework to explain the origin of a non-universal, bottom-heavy, IMF in massive galaxies. Indeed, a top-heavy IMF has been invoked to explain the high $\rm [\alpha/Fe]$ observed in massive ETGs, since the downsizing in star-formation alone is not able to reproduce the values of $\rm [\alpha/Fe]$ observed (De Masi et al., in prep.). Furthermore, the possibility of an IMF slope dependent on the instantaneous star formation rate (SFR) has been proposed by \citet{2011MNRAS.415.1647G} and \citet{2013MNRAS.435.2274W}, with the IMF becoming increasingly top-heavy with increasing SFR. This idea has also been tested in the GAEA semi-analytic models \citep{2017MNRAS.464.3812F,2017MNRAS.466L..88D}, where an IMF changing with the instantaneous SFR has been found to reproduce the enhanced $\rm [\alpha/Fe]$ of ETGs. In order to reconcile the high metal content and enhanced $\rm [\alpha/Fe]$ of ETGs with a bottom-heavy IMF, a time-dependent scenario seems to be actually required, where the IMF switches from a top- to bottom-heavy phase during the initial phases of collapse \citep{1997ApJS..111..203V,2013MNRAS.435.2274W,2015MNRAS.448L..82F}, perhaps due to the rapid injection of energy into a highly dense and turbulent interstellar medium (\citealt{2013MNRAS.433..170H}; see also \citealt{2014ApJ...796...75C}). The mechanism(s) behind such variations of IMF slope with SFR and time are still unclear, though. In this regard, studying the dependence of IMF on environment might provide some further clue, as galaxies belonging to different environments are expected to have experienced different physical conditions at their formation, as well as different SFHs during their evolution.\par
Last, but not least, radial gradients in IMF slope have been recently found for a number of massive ETGs \citep{2015MNRAS.447.1033M,2016MNRAS.457.1468L}. The central regions of these objects show a bottom-heavy IMF, while the outskirts follow a Kroupa-like distribution (but see \citealt{2017MNRAS.468.1594A}). Interestingly, this feature could be explained in light of the two-phase formation scenario for massive ETGs described above.\par
In the  present work, we focus on the spectroscopic approach to constrain the IMF, measuring, for the first time, how the IMF changes with velocity dispersion in ETGs, as a function of hierarchy as well as the environment where galaxies reside. To this aim, we analyse a set of optical and NIR line-strengths in stacked spectra of ETGs from the Sloan Digital Sky Survey (SDSS), following a similar methodology as that adopted in our previous works \citep{2013MNRAS.429L..15F,2013MNRAS.433.3017L,2015MNRAS.449L.137L}. The layout of the paper is the following. In Sections~\ref{sec:data} and \ref{sec:ssps} we present the data and models used in the analysis, respectively. Section~\ref{sec:meas} describes our methodology, while Section~\ref{sec:res} and \ref{sec:disc} present and discuss the results. Conclusions are drawn in Section~\ref{sec:conc}.
\section{Data}
\label{sec:data}
The SPIDER\footnote{Spheroids Panchromatic  Investigation in Different Environmental Regions, \citet{2010MNRAS.408.1313L}.} catalogue contains 39993 galaxies in the redshift range $0.05<z<0.095$, classified as ETGs because of their passive spectra and bulge dominated morphology (following the definition of \citealt{2003AJ....125.1849B}). The bona-fide ETGs version of the catalogue, with better quality SDSS spectroscopy available (see \citealt{2013MNRAS.433.3017L}), also imposes that the galaxies meet the following criteria:
\begin{itemize}
\item central velocity dispersion $\sigma_0 \ge 100$ km/s,
\item $\rm E(B-V)<0.1$ mag,
\item $\rm S/N(\text{\AA}^{-1})> (14,27,21)$ for $\sigma_0=(100,200,300)$ km/s,
\end{itemize}
which results in a reduced sample of $\rm N_G=24781$ SPIDER ETGs. Finally, a visual inspection of the morphology of these objects, aimed at removing late-type galaxies with a prominent bulge, reduces the sample to 21665 ETGs.\par
We match the final SPIDER ETG sample with the 596851 galaxies in the group catalogue by \citet{2014MNRAS.439..611W}, which is the version of the group catalogue of \citet{2007ApJ...671..153Y} updated to SDSS DR7, and obtain a final sample of 20996 SPIDER ETGs for which we have both environmental information from the group catalogue and a spectrum available from SDSS (DR12).\par
We correct the flux of the retrieved spectra for Galactic extinction, using the Schlegel maps obtained from the IRSA website\footnote{http://irsa.ipac.caltech.edu/applications/DUST/} and by adopting the Galactic extinction law by \citet{1989ApJ...345..245C}. We also correct the spectra for redshift and transform them from the vacuum system to the air system, following \citet{1991ApJS...77..119M}, in order to later compare them with MIUSCAT/EMILES synthetic stellar population models (see Sect.~\ref{sec:ssps}). To this effect, the spectra were interpolated with a linear spline into a common wavelength grid, spanning the range 3800-8800 {\AA}, with a fixed dispersion of 1 {\AA}. We chose the wavelength range to be in common to most of the spectra, at the same time including all the (optical and NIR) absorption features required for the analysis (see \citealt{2013MNRAS.433.3017L} for details). Finally, we redefined the uncertainty on the flux by reinterpolating the spectrum, offset by $\rm \pm 1\sigma_{flux}$, respectively, and then by taking the halved difference between the two interpolated spectra as the new uncertainty.
\subsection{Environment}
\label{sec:envir}
\begin{figure*}
\hspace*{-0.6cm}
\includegraphics[width=130mm]{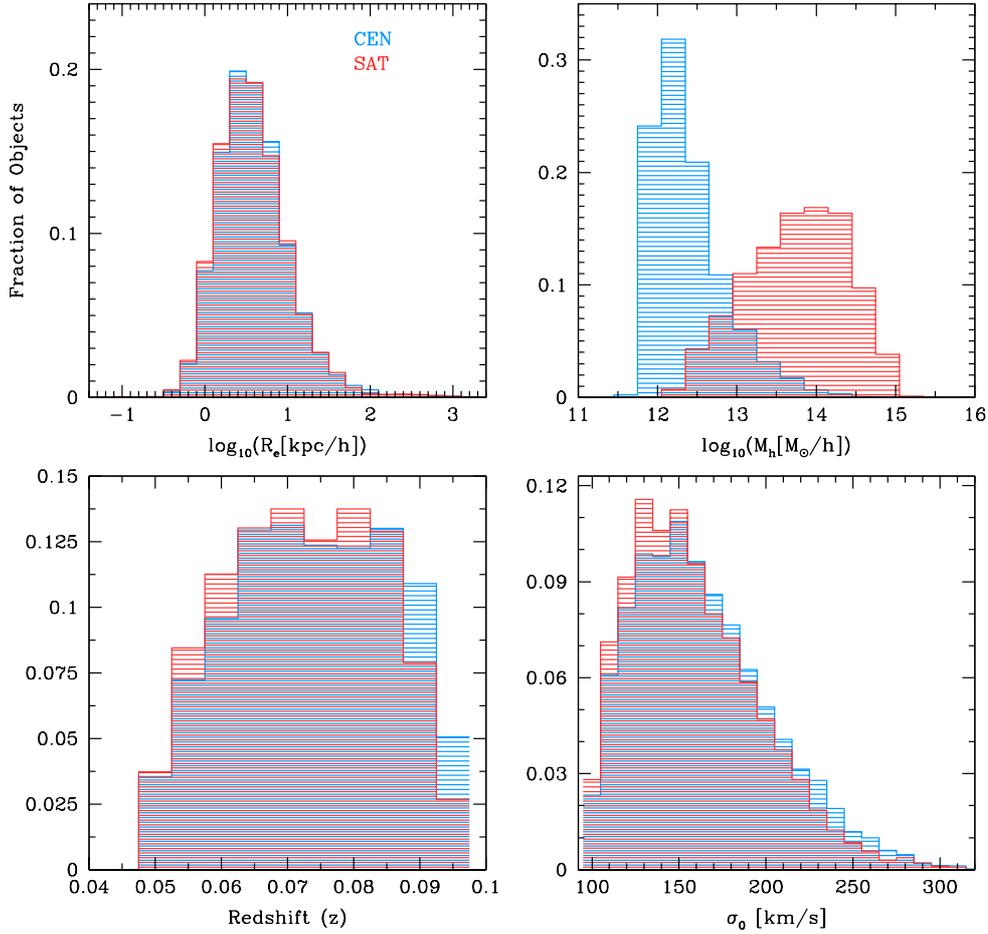}
\caption{Distribution of various properties for centrals (in blue) and satellites (in red). The upper two panels show respectively the distribution of the sample galaxies in effective radius (left) and host halo mass (right), whereas the lower two panels show the distribution in redshift (left) and central velocity dispersion (right). There is only a distinction in distribution between centrals and satellites when they are binned in host halo mass.}
\label{fig:isto}
\end{figure*}
The \citet{2014MNRAS.439..611W} and the \citet{2007ApJ...671..153Y} catalogues use an iterative routine to find galaxy groups and assign a galaxy to a given group, which is represented by its dark matter host halo mass. The routine first uses an FOF algorithm with small linking lengths in redshift space to tentatively assign galaxies to groups and estimate the group's total stellar mass through its total luminosity. It then uses an iterative procedure to assign a mass to the dark matter host halo of the group based on the average M/L of the groups found in the previous iteration. With a mass assigned to the host halo, the routine estimates the size and velocity dispersion of the group and reassigns the membership. This part is repeated until convergence to a final result is reached. The routine also labels the most massive galaxy in the group as the central, while all the other galaxies assigned to the group are considered satellites. The calculation of the stellar mass of the galaxies is performed using the relation of \citet{2003ApJS..149..289B} between stellar mass-to-light ratio and colour. The dark matter masses, on the other hand, have been obtained using both the total characteristic luminosity and stellar mass of the groups (see \citealt{2007ApJ...671..153Y} for more details).\par
Hence, for each SPIDER ETG, the information we extract from the \citet{2014MNRAS.439..611W} catalogue is the dark matter halo mass of its parent group, as derived from the group total stellar mass~\footnote{As shown in \citet{2011MNRAS.410..210M}, the total stellar mass is a better proxy of the halo mass compared to total luminosity.}, and the hierarchy of the galaxy itself (satellite or central). We first split the galaxy sample into two subsamples, based on galaxy hierarchy: subsample CEN for the centrals ($15571$ objects), and subsample SAT for the satellites ($5425$ objects). This first subdivision is made regardless of the mass of the host halo the galaxies reside in and allows us to see if galaxy hierarchy influences the properties derived from the spectra, i.e. age, metallicity, $\rm [\alpha/Fe]$, and in particular, the IMF slope ($\rm \Gamma_b$).\par
In addition, we created two subsamples for each hierachical subsample by differentiating between galaxies inhabiting high and low mass host haloes. The cut in host halo mass was set to $\rm 10^{12.5}\, M_\odot/h$ and $\rm 10^{14}\, M_\odot/h$ for centrals and satellites, respectively, resulting in the following subsamples:
\begin{itemize}
\item C1 ($\rm M_h < 10^{12.5}\, M_\odot/h$) with $\rm N_{ETGs}=$10515;
\item C2 ($\rm M_h \ge 10^{12.5}\, M_\odot/h$) with $\rm N_{ETGs}=$5027;
\item S1 ($\rm M_h < 10^{14}\, M_\odot/h$) with $\rm N_{ETGs}=$3284;
\item S2 ($\rm M_h \ge 10^{14}\, M_\odot/h$) with $\rm N_{ETGs}=$2093.
\end{itemize}
This division closely follows the one made in \citet{2014MNRAS.445.1977L}, hereafter LB14, with the only difference being that the satellite subsample is not further subdivided with respect to their group-centric distance. The difference in the mass cut is justified by the distribution of host halo mass for centrals and satellites, as seen in Fig.~\ref{fig:isto} (top--right). Furthermore, the cut at $\rm M_h=10^{12.5}\, M_\odot/h$ for centrals is compatible with a halo hosting an $\rm L^*$ galaxy \citep{2010ApJ...710..903M}.\par
Fig.~\ref{fig:isto} shows the distribution of four properties of our galaxy sample for both the CEN and the SAT subsamples. We see that the histograms of centrals and satellites are very similar, except for their distribution in host halo mass. This is not surprising, since lower mass haloes are believed to be more frequent in the Universe and since centrals follow the host halo distribution closely. On the other hand, the distribution of satellites results from the fact that these galaxies are more numerous in more massive host haloes, but at the same time such haloes are rare in the Universe. Hence, following LB14, we adopt two different host halo mass cuts for centrals and satellites, based on the peaks of the CEN and SAT distributions. Finally, we notice that the number of objects given above for subsamples C1/C2 and S1/S2 does not sum to the number of objects given for CEN and SAT, because the subsamples C1/C2 and S1/S2 are counted after the bins in central velocity dispersion have been constructed, and some galaxies have been rejected accordingly (see Sec.~\ref{sec:stack} for details). After such binning the CEN and SAT subsamples are reduced to 15559 and 5408 objects, respectively.
\subsection{Stacked spectra}
\label{sec:stack}
To measure the effect of the IMF on absorption features, we need high signal-to-noise ratio spectroscopy ($\rm S/N \gtrsim 100$ {\AA}$^{-1}$; see, e.g., \citealt{2012ApJ...747...69C}). To this aim, we stack the spectra of ETGs in central velocity dispersion ($\sigma_0$) bins, following a similar procedure as in \citet{2013MNRAS.433.3017L}, hereafter LB13, and LB14. The $\sigma_0$-bins span the range $[100,310]$ km/s, and are generally $10$ km/s wide. They are defined so as to contain at least $40$ objects each, and, should this not be the case, they are widened by $10$ km/s until a maximum width of $30$ km/s is reached. If the condition is still not met, the bin is rejected. The stacking procedure allows us to raise the quality of the spectra, but at the cost of obtaining an average behaved spectrum over the galaxy population in the bin.\par
For each bin, in order to account for differences in the absorption features due to the galaxies' kinematics, we broaden the spectra to the upper value of $\sigma_0$ in the bin, by convolving the spectra with an appropriate gaussian function. We then proceed to stack the spectra, by first normalising each of them by its median flux in the wavelength range $[5000-8000]$ {\AA}. Once this is done for all the spectra in the bin, we multiply all normalised spectra by the median flux of the median fluxes found for each individual spectrum. This puts all the spectra in one bin at roughly the same flux level and allows differences in flux to be exclusively due to noise.\par
We compute the stacked spectrum by taking the median flux, at each wavelength, of all the processed spectra in the $\sigma_0$ bin. The resulting stacks have an enhanced signal-to-noise ratio, by a factor of at least $\sim 10$, with respect to the single spectra in each bin.\par
\begin{figure}
\hspace*{-0.6cm}
\includegraphics[width=90mm]{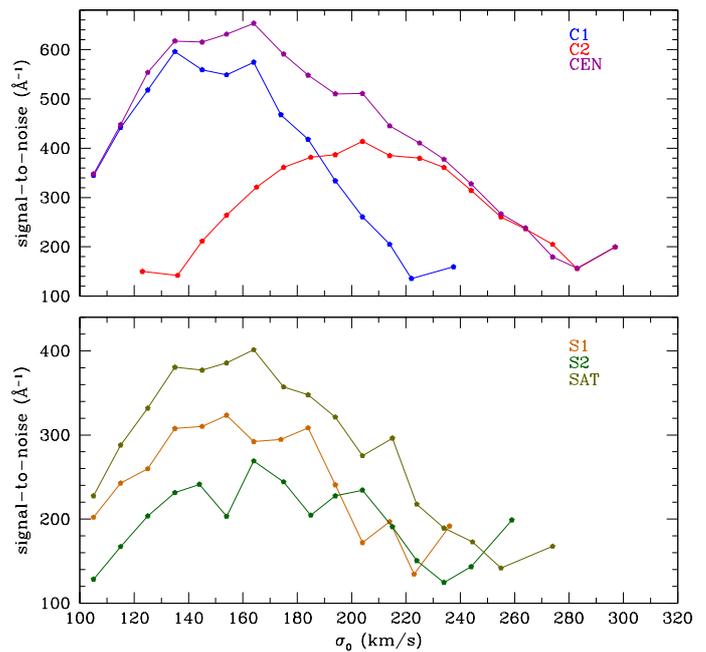}
\caption{S/N  ratio of  the stacked  spectra analyzed  in the  present work, as a function of $\sigma_0$. The upper and lower panels refer to the different subsamples of central and satellite ETGs, respectively.}
\label{fig:SNcomp}
\end{figure}

Fig.~\ref{fig:SNcomp} shows the median S/N ratio of the stacked  spectra -- measured per {\AA} in the region $[4840-4880]$ {\AA} -- as a function of $\sigma_0$, for our different galaxy subsamples. All stacks have a S/N above $100$. During the procedure the resolution of the spectra remained that of SDSS, while the dispersion was fixed at 1 {\AA}/pix for the whole wavelength range.
\section{Synthetic Stellar Population Models}
\label{sec:ssps}
For each stacked spectrum we measure the line strengths for a given set of spectral indices and compare them to those predicted from synthetic stellar population models. The models used in this work are the MIUSCAT models of \citet{2012MNRAS.424..157V}, and the EMILES models of \citet{2016MNRAS.463.3409V}.\par
The MIUSCAT models cover the wavelength range $[3465-9469]$ {\AA} and are constructed using the MILES \citep{2006MNRAS.371..703S}, CaT \citep{2001MNRAS.326..959C} and Indo-U.S. \citep{2004ApJS..152..251V} empirical stellar libraries. These libraries cover the wavelength range $[3525-7500]$ {\AA}, $[8350-9020]$ {\AA}, and $[3465-9469]$ {\AA}, respectively. The Indo-U.S. is only used to fill the gap between the MILES and CaT spectral libraries. MIUSCAT models are computed at a fixed spectral resolution of $2.51$ {\AA} (FWHM).\par
EMILES models extend MIUSCAT both bluewards and redwards, to $1680.2$ {\AA} and $\sim  5\,  \mu$m, respectively. These models are constructed using spectra from the IRTF\footnote{Infrared Telescope Facility} stellar library \citep{2005ApJ...623.1115C,2009ApJS..185..289R} to extend the MIUSCAT models towards the infrared \citep{2016A&A...589A..73R}, and using spectra from the NGSL\footnote{Next Generation Spectral Library} spectral library \citep{2006hstc.conf..209G,2012A&A...538A.143K} to extend MIUSCAT bluewards.\par
Both MIUSCAT and EMILES SSPs are computed for several IMFs, including unimodal (single power-law) and bimodal (low-mass tapered) IMFs, both characterized by their slope, $\rm \Gamma$ (unimodal) and $\rm \Gamma_b$ (bimodal), as a single free parameter (see, e.g.,~\citealt{1996ApJS..106..307V,2003MNRAS.340.1317V,2015MNRAS.448L..82F}). The bimodal IMFs are given by a power-law smoothly tapered off below a characteristic ``turnover'' mass of $0.6$\,M$_\odot$. For $\rm \Gamma_b \sim 1.3$, the bimodal IMF gives a good representation of the Kroupa IMF, while for $\rm \Gamma\sim 1.35$ the unimodal IMF coincides with the \citet{1955ApJ...121..161S} distribution. The lower and upper mass-cutoff of the IMFs are set to $0.1$ and $100$\,M$_\odot$, respectively. Since the bimodal distribution consists of a power-law at the high-mass end, while it is smoothly tapered towards low masses, varying $\rm \Gamma_b$ changes the dwarf-to-giant ratio in the IMF through the normalization. While this approach is different with respect to a change of the IMF slopes at low- and very-low mass (e.g.~\citealt{2012ApJ...760...71C}), this parameterisation is good enough for our purposes, as in the present work we do not aim at constraining the IMF shape in detail, but rather to study the possible dependence of IMF variations on galaxy environment. In addition, the bimodal IMF has been found to provide a consistent explanation between optical and NIR IMF-sensitive features, and consistent constraints to dynamical models, in contrast to the unimodal distribution \citep{2016MNRAS.457.1468L,2016MNRAS.463.3220L}.\par
Since the wavelength range of our stacked spectra is fully covered by MIUSCAT/EMILES models, for the purpose of our analysis, the only difference between the two sets of models is the range of IMF slope for which they are computed. EMILES reaches a higher maximum value of $\rm \Gamma_b=3.5$ (instead of $3.3$, for MIUSCAT), with a better sampling of the range $\rm 2.0<\Gamma_b<3.0$ and thus we compare the stacked spectra to EMILES models in our final results.\par
Both MIUSCAT and EMILES models have been generated using the isochrones of \citet{2000A&AS..141..371G}, hereafter called Padova isochrones. In addition to the Padova isochrones, we also generated EMILES models using the isochrones of \citet{2004ApJ...612..168P} and \citet{2006ApJ...642..797P}, hereafter Teramo isochrones. The MIUSCAT and EMILES Teramo models were used to test the robustness of our results against the ingredients of stellar population models.\par
Table~\ref{table:mod} summarizes the values of $\rm \Gamma_b$, metallicity and age used to construct the models. Different sets of isochrones have different values of age and metallicity, with a different sampling. While the Padova isochrones sample their age range logarithmically with smaller steps for younger ages with respect to older ages, the Teramo isochrones sample the age range linearly with a step that varies based on the age of the stellar population (see \citealt{2015MNRAS.449.1177V} for details). In Section~~\ref{sec:diffmod}, we compare results obtained for different models.
\begin{table*}
\centering
\begin{tabular}{|c|c|c|c|}
\hline
Model&bimodal IMF Slope ($\rm \Gamma_b$)&Metallicity [$\rm Z/H$]&Age (Gyr)\\
\hline
&0.30, 0.80, 1.00,&-1.71, -1.31,&0.063\\
MIUSCAT& 1.30, 1.50, 1.80, 2.00,&-0.71, -0.40,&-\\
(Padova00)&2.30, 2.80, 3.30&+0.00, +0.22, +0.40&17.7828\\
\hline
&0.30, 0.50, 0.80, 1.00,&-2.32, -1.71, -1.31,&0.0631\\
EMILES&1.30, 1.50, 1.80, 2.00,&-0.71, -0.40&-\\
(Padova00)&2.30, 2.50, 2.80, 3.00, 3.30, 3.50&+0.00, +0.22&17.7828\\
\hline
&0.30, 0.50, 0.80, 1.00,&-2.27, -1.79, -1.49, -1.26,&0.0300\\
EMILES& 1.30, 1.50, 1.80, 2.00,& -0.96, -0.66, -0.35, -0.25&-\\
(Teramo)&2.30, 2.50, 2.80, 3.00, 3.30, 3.50&+0.06, +0.15, +0.26, +0.40&14.0000\\
\hline
\end{tabular}
\caption{The table shows the IMF slopes, metallicities and ages used to create the synthetic spectra of the three sets of simple stellar population models used in our analysis.}
\label{table:mod}
\end{table*}
\section{Analysis of line strengths}
\label{sec:meas}
In order to constrain galaxy age, metallicity, $\rm [\alpha/Fe]$, and IMF slope, for each stacked spectrum we measure the equivalent widths of a set of specific line indices. The age parameter is mainly constrained through $\rm  H\beta_o$, the optimized $\rm H\beta$ index defined by \citet{2009MNRAS.392..691C}. To this effect, we correct the index for nebular emission contamination with a similar procedure as that described in LB13, i.e. estimating the excess of flux in the line with respect to a combination of two MIUSCAT SSPs that best fit the \hb\ spectral region ($\lambda\lambda = 4530-4730$\,\AA) when excluding the trough of the absorption. For each stack, the emission correction is determined separately for models with different $\rm \Gamma_b$, and then applied iteratively when the IMF slope is being determined (see below). Metallicity is mainly constrained through the total metallicity indicator \mgfep~\citep{2003MNRAS.339..897T}, which is insensitive to $\rm [\alpha/Fe]$.\par
Table~\ref{table:ind} shows the set of line indices used in the fits (the two different cases, ``.0'' and ``.1'', will be explained in Subsection~\ref{sec:fit}) and the expected sensitivity of each index to different stellar population properties. We also report the references where the central passband and pseudo-continuum bands of the lines have been defined. Additionally, Appendix~\ref{sec:ind} shows the trend of all spectral indices, for all stacked spectra, as a function of central velocity dispersion $\sigma_0$.
\begin{table*}
\centering
\begin{tabular}{c c c}
\hline
Method&Lines used&Abundances\\
\hline
``.0''&$\rm H\beta_o$, $\rm [MgFe]'$, $\rm TiO1$, $\rm TiO2_{SDSS}$, $\rm Mg4780$,&None\\
& $\rm NaI8190$, $\rm Ca2$, $\rm CaH\&K$, $\rm NaD$&\\
&&\\
``.1''&$\rm H\beta_o$, $\rm [MgFe]'$, $\rm TiO1$, $\rm TiO2_{SDSS}$, $\rm Mg4780$,&$\rm [Ca/Fe]$, $\rm [Na/Fe]$, $\rm [Ti/Fe]$,\\
& $\rm NaI8190$, $\rm Ca2$, $\rm CaH\&K$, $\rm NaD$,& $\rm [O/Fe]$, $\rm [C/Fe]$, $\rm [N/Fe]$,\\
& $\rm Ca1$, $\rm Ca4227$, $\rm Fe4531$, $\rm Mg1$,& $\rm [Mg/Fe]$, $\rm [Si/Fe]$\\
& $\rm Mg2$, $\rm C4668$, $\rm CN2$, $\rm Mgb5177$&\\
\hline
\hline
\end{tabular}
\footnotesize
\begin{tabular}{c c c c c}
&&&\\
Index&IMF sensitive&Abundance Fit&Other Constraint&Definition\\
\hline
$\rm CaH\&K$&Yes&$\rm [Ca/Fe]$&/&\citet{2005ApJ...627..754S}\\
$\rm CN2$&No&$\rm [C/Fe]$,$\rm [N/Fe]$,$\rm [O/Fe]$&/&\citet{1998ApJS..116....1T}\\
$\rm Ca4227$&No&$\rm [Ca/Fe]$&/&\citet{1998ApJS..116....1T}\\
$\rm Fe4531$&No&$\rm [Ti/Fe]$&/&\citet{1998ApJS..116....1T}\\
$\rm C4668$&No&$\rm [C/Fe]$&/&\citet{1998ApJS..116....1T}\\
$\rm Mg4780$&Yes&No&/&\citet{2005ApJ...627..754S}\\
$\rm H\beta_o$&No&No&Age Indicator&\citet{2009MNRAS.392..691C}\\
$\rm Mg1$&No&$\rm [C/Fe]$,$\rm [O/Fe]$,$\rm [Si/Fe]$&/&\citet{1998ApJS..116....1T}\\
$\rm Mg2$&No&$\rm [Mg/Fe]$,$\rm [Si/Fe]$&/&\citet{1998ApJS..116....1T}\\
$\rm Mgb5177$&No&$\rm [Mg/Fe]$&[$\rm \alpha/Fe$] Proxy&\citet{1998ApJS..116....1T}\\
$\rm [MgFe]'$&No&No&Z-metallicity Indicator&\citet{2003MNRAS.339..897T}\\
$\rm Fe3$&No&No&[$\rm \alpha/Fe$] Proxy&\citet{2000MNRAS.315..184K}\\
$\rm NaD$&Yes&$\rm [Na/Fe]$&/&\citet{1998ApJS..116....1T}\\
$\rm TiO1$&Yes&$\rm [Ti/Fe]$&/&\citet{1998ApJS..116....1T}\\
$\rm TiO2_{SDSS}$&Yes&$\rm [Ti/Fe]$&/&\citet{2013MNRAS.433.3017L}\\
$\rm NaI8190$&Yes&$\rm [Na/Fe]$&/&\citet{2013MNRAS.433.3017L}\\
$\rm Ca1$&No&$\rm [Ca/Fe]$&/&\citet{2001MNRAS.326..959C}\\
$\rm Ca2$&Yes&No&/&\citet{2001MNRAS.326..959C}\\
\end{tabular}
\caption{Different sets of line indices used in the ``.0'' and ``.1'' methods to fit the observed line-strengths of stacked spectra to the ones obtained from SSP models (see text) when determining the IMF. The upper table also shows which abundances are fitted in the ``.1'' case and is analogous to the table in \citet{2015MNRAS.449L.137L}, where a similar approach has been adopted. The lower part of the table shows how different line indices are defined and have been used in the fits, i.e. if they constrain the IMF, one or more of the given abundance ratios, age, metallicity, and $\rm [Mg/Fe]$. References, reporting where the feature and pseudo-continuum bands of each index are defined, are listed in the last column of the lower table.}
\label{table:ind}
\end{table*}

\subsection{Fitting the measured indices}
\label{sec:fit}

We compare observed line-strengths to a grid of predictions for SSP models with varying age, metallicity, and IMF slope. The grids are constructed by linearly interpolating the models performing $200$ and $150$ steps in age and metallicity, respectively. For each stacked spectrum, model line-strengths are computed by first smoothing both MIUSCAT and EMILES SSPs to match the $\sigma_0$ of the given stack. We also take the effect of instrumental resolution into account and its dependence on wavelength when smoothing the models to match the observed spectra. We consider two fitting approaches, indicated as ``.0'' and ``.1'', respectively.\par
In the case ``.0'', we adopt an approach very similar to that of LB13. We fit \hbo~and \mgfep\ (to  constrain  age  and   metallicity), plus a number of IMF-sensitive features, i.e. $\rm NaI8190$, \caii~\footnote{The main difference with respect to LB13 is that we do not consider the combined calcium-triplet index (CaT) in the present analysis, but only the \caii. The reason for this choice is that the third CaT line, \caiii, is at the border of the SDSS spectral range, where the low quality of the spectra is not sufficient for our purposes, when binning spectra as function of both $\sigma_0$ (as in LB13) and environment/hierarchy. For the same reason, we include only \cai\ and \caii\ in the ``.1'' fitting approach (see below). \cai\ is excluded from the ``.0'' case, because it is more sensitive to $\rm [Ca/Fe]$ than IMF variations if compared to \caii.}, \tioi, \tioiio,  and \mgf, as well as~\nad\ and \cahk\ (which have some sensitivity to IMF, as well as to abundance ratios). We do {\it not} fit individual abundance ratios, but compare directly observed and model line-strengths. Since we rely on MIUSCAT/EMILES models, the abundance patterns, $\rm [X/Fe]$ (where $\rm X$ denotes a generic element), of the models follow closely those of stars in the disk of our Galaxy, i.e. they are approximately solar-scaled at solar and super-solar metallicity. On the contrary, massive ETGs have non-solar abundance ratios (\citealt{1989PhDT.......149P,1992ApJ...398...69W,1995A&A...296...73W}, but see also \citealt{2006ARA&A..44..141R} and references therein). To take this into account, we  first correct all observed line-strengths to solar scale, following the method in LB13. In practice, we estimate $\rm [Mg/Fe]$ (independently of the ``.0'' case) for each stacked spectrum, by fitting the $\rm Mgb5177$ and $\rm Fe3$ indices, at fixed age (estimated through the \hbo--\mgfep\ diagram), with models that have a varying metallicity at fixed IMF slope~\footnote{We adopt an IMF slope of \gammab$=1.3$, although the estimate of \mgfe\ does not depend significantly on \gammab, as shown in \citet{2017MNRAS.464.3597L}.}. The fits provide two metallicity estimates, $\rm [Z/H]_{Mg}$ and $\rm [Z/H]_{Fe}$, respectively. We derive \mgfe\ from the following ansatz:
\begin{equation}
\begin{array}{rl}
\rm [Mg/Fe]&\rm =0.55[Z_{Mg}/Z_{Fe}]\\
&\rm =0.55([Z/H]_{Mg}-[Z/H]_{Fe}),\\
\end{array}
\label{eq:MgFe}
\end{equation}
where the coefficient $0.55$ has been determined in LB13, based on model predictions from \citet{2011MNRAS.412.2183T} (see also \citealt{2015MNRAS.449.1177V}). The estimate of \mgfe\ is then used to correct observed line-strengths to solar-scale (see LB13 for details).\par
The ``.0'' fitting procedure is performed by minimizing the $\chi^2$:
\begin{equation}
\chi^2(\Gamma_b)= \sum_{indices} \frac{(I_{measured}^{corr}-I_{mod})^2}{\sigma_I^2+s_I^2},
\label{eq:fit}
\end{equation}
where $I_{measured}^{corr}$ are the equivalent widths of the stacked spectra, corrected to solar scale; $I_{mod}$ are the model line-strengths; $\sigma_I$ and $s_I$ are the uncertainties on the measured line-strengths, and on the correction for non-solar abundance pattern, respectively. For $\rm H\beta_o$, the correction for nebular emission is also applied iteratively~\footnote{In practice, we start by applying the emission correction obtained with models having a Kroupa-like IMF. Then, once the IMF slope is derived, we apply the emission correction computed for that IMF, and repeat the whole fitting procedure. In general, the second estimate of IMF slope is very similar to that obtained in the first step, without any need of a further iteration.}. The parameters fitted in the ``.0'' case are age, metallicity and the IMF slope $\rm \Gamma_b$.\par
In the second fitting approach, named ``.1'', the $I_{mod}$ terms in Eq.~\ref{eq:fit} are redefined as $I_{mod.1}=I_{mod}+\delta_{X} \rm [X/Fe]$, where $\delta_{X}=\delta(I)/\delta(\rm [X/Fe])$ is the sensitivity of a given index $\rm I$ to a variation in the abundance pattern of element $\rm X$. The coefficients $\rm \delta_{X}$ are computed with the aid of \citet{2012ApJ...747...69C} stellar population models (CvD12 models), having a Chabri\'er IMF, old age ($12.5$~Gyr), and solar metallicity. The $\rm [X/Fe]$ abundance ratios are treated as free fitting parameters, together with IMF slope, age, and (total) metallicity. In order to constrain $\rm [X/Fe]$ properly, we enlarge the set of targeted spectral features with respect to case ``.0'', including also indices that are sensitive to individual abundance ratios. The list of indices used in both fitting aproaches are summarized in Table~\ref{table:ind}. Comparing the results of cases ``.0''  and ``.1'', we test the robustness of our results, and investigate if and how much they are affected by possible degeneracies between IMF and abundance ratios. For case ``.1'', the fits do also provide abundance ratio estimates for our stacked spectra. However, studying the dependence of $\rm [X/Fe]$ on $\sigma_0$, environment, and hierarchy, is beyond the scope of this paper, and will be eventually presented in a forthcoming work.

\section{Results}
\label{sec:res}

We present a comparison of age, metallicity, $\rm [Mg/Fe]$, and IMF slope for the different subsamples of centrals (Sect.~\ref{sec:c12}), satellites (Sect.~\ref{sec:s12}), and finally as a function of galaxy hierarchy (Sect.~\ref{sec:cs}). We base our main results on EMILES Padova models, as different models (i.e. MIUSCAT vs. EMILES; as well as different sets of isochrones) and different assumption on SFHs (2SSP vs 1SSP) give very consistent results, as shown in Sec.~\ref{sec:diffmod}.

\subsection{Comparing centrals}
\label{sec:c12}
\begin{figure}
\hspace*{-0.5cm} \includegraphics[width=90mm]{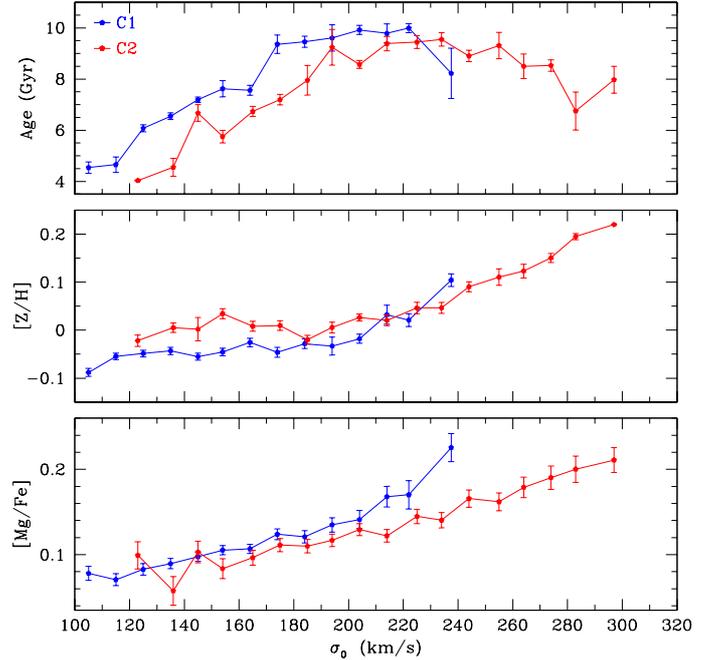}
\caption{Trend of age, metallicity and $\rm [Mg/Fe]$ with $\sigma_0$ for centrals in low- and high-mass haloes (subsamples C1 and C2), respectively. Age and metallicity are estimated through method ``.0'', while $\rm [Mg/Fe]$ is obtained from the ``.1'' fitting approach (see the text for details). C2 ETGs exhibit younger ages, higher metallicity and lower $\rm [Mg/Fe]$, than those in the C1 subsample, in agreement with \citet{2014MNRAS.445.1977L}.}
\label{fig:prop1}
\end{figure}

Fig.~\ref{fig:prop1} compares age, metallicity and $\rm [Mg/Fe]$ -- which is a proxy of $\rm [\alpha/Fe]$ -- for our subsamples of centrals, residing in low- (C1) and high- (C2) mass haloes, respectively. We consider metallicity and age estimates from method ``.0'' only. The reason for this choice is the following. In method ``.1'' we also include abundance ratios in the fitting procedure, computing the sensitivity of different indices to elemental abundances with CvD12 models (see above). Such models are computed at fixed $\rm [Fe/H]$, rather than total, metallicity (i.e. MIUSCAT/EMILES). Therefore, we consider the metallicity estimate from method ``.1'' less reliable than that from method ``.0''. For $\rm [Mg/Fe]$, instead, we consider results from method ``.1'', as $\rm [Mg/Fe]$ is not fitted in the ``.0'' case. Notice that this approach is different from that of LB14, where we used  the solar-scale proxy for $\rm [Mg/Fe]$ (see Eq.~\ref{eq:MgFe}) and age/metallicity estimates inferred from the \hbo--\mgfep\ diagram. Our current results are very similar to those of LB14, though, as shown below.\par
We see that C2 centrals in high mass haloes are generally slightly younger and metal richer (at $\sigma_0 \lesssim 220$~km/s) and have a slightly lower $\rm [Mg/Fe]$ with respect to C1 centrals in low-mass host haloes. We find an average difference of $\Delta Age=1.2$~Gyr, $\Delta \rm [Z/H]=0.041$~dex and $\Delta \rm [Mg/Fe]=0.03$~dex, corresponding to a significance level of $\sim 4 \sigma$, $\sim 2.6 \sigma$, $\sim 1.7 \sigma$ respectively. The metallicity and the $\rm [Mg/Fe]$ ratio rise with increasing $\sigma_0$, while the age first rises and then flattens around $\sigma_0=200$ km/s.\par
Fig.~\ref{fig:prop1} is directly comparable to Fig.~$3$ of LB14, where similar trends of age, metallicity and $\rm [Mg/Fe]$ with $\sigma_0$ and halo mass are shown. We notice, however, that given the difference in methods used to derive these properties, our values of $\rm [Z/H]$ and $\rm [Mg/Fe]$ have an offset with respect to LB14 of $+0.1$~dex and $-0.05$~dex, respectively. Furthermore, the derived ages are lower by about $2$~Gyr for the lowest and highest $\sigma_0$ bins.\par
\begin{figure}
\hspace*{-1cm}
\includegraphics[width=97mm]{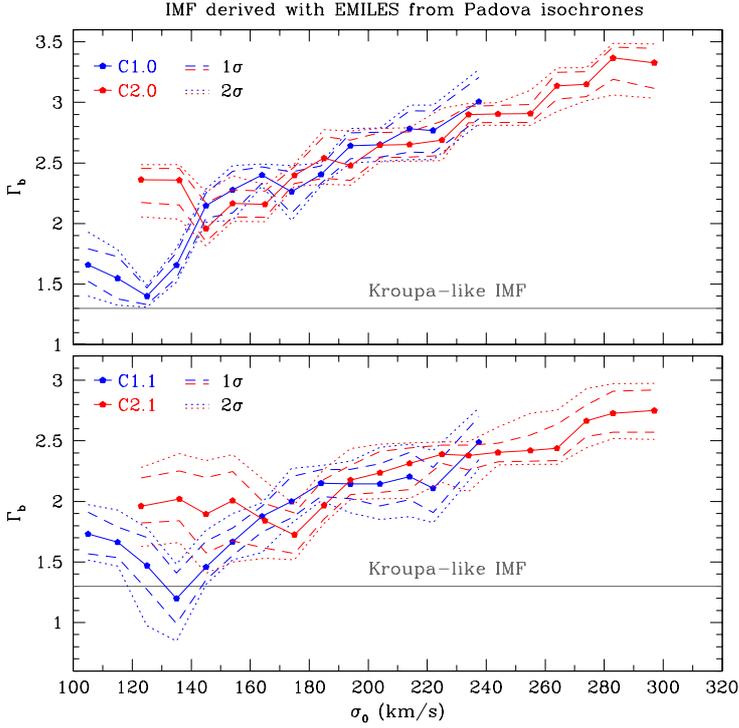}
\caption{IMF slope, estimated through EMILES models with Padova isochrones \citep{2016MNRAS.463.3409V}, for different subsamples of central ETGs. The upper panel shows the results of fitting only IMF sensitive indices (plus \hbo\ and \mgfep; i.e. the case ``.0''), while the lower panel shows the results obtained when abundance ratios are also fitted (case ``.1''). The dashed and dotted lines represent the $1\sigma$ and $2\sigma$ contours, respectively. We see no influence of environment on IMF slope for centrals, except for the two lowest bins of $\sigma_0$, where C2 ETGs have systematically higher \gammab. This issue is further investigated in Sect.~\ref{sec:probC12}.}
\label{fig:myEC12}
\end{figure}

The IMF slopes of subsamples C1 and C2, for both the ``.0'' and ``.1'' cases, are shown in Fig.~\ref{fig:myEC12}. We see that the value of $\rm \Gamma_b$ also rises with increasing $\sigma_0$, turning the IMF from a Kroupa-like function (\gammab~$\sim 1.3$) to a bottom-heavy distribution (\gammab~$\gtrsim 2.5$) at high central velocity dispersion. We do not observe a trend with host halo mass in the comparison of these two subsamples, but we notice that, for the lowest values of $\sigma_0$, C1 and C2 significantly differ in IMF slope. This issue is further investigated and discussed in Section~\ref{sec:probC12}.
\subsection{Comparing satellites}
\label{sec:s12}
\begin{figure}
\hspace*{-0.5cm}
\includegraphics[width=90mm]{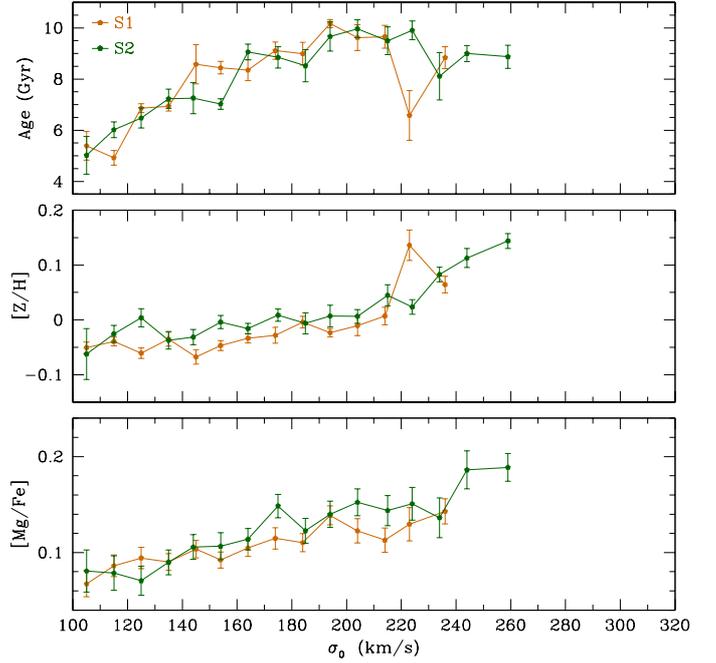}
\caption{Same as Fig.~\ref{fig:prop1}, but for the S1 and S2 subsamples.}
\label{fig:prop3}
\end{figure}

Fig.~\ref{fig:prop3} shows the comparison of age, metallicity, and $\rm [Mg/Fe]$, for our satellite subsamples, S1 and S2, respectively. All galaxy properties show the same behaviour with $\sigma_0$ as for the central subsamples (C1 and C2), with age, metallicity, and $\rm [Mg/Fe]$ all increasing with galaxy velocity dispersion. Differently from centrals, the trend of age with $\sigma_0$ for satellites does not depend significantly on environment (i.e. host halo mass), within the corresponding uncertainties (i.e. the average age difference is significant only at the $\sim 1.3\sigma$ level). The same can be said for the differences in metallicity and $\rm [Mg/Fe]$, which are only significant at $\sim 1.5\sigma$ and $0.8\sigma$, respectively. We conclude that environment does not influence the average properties of our satellite subsample, to our current precision.\par
Fig.~\ref{fig:prop3} is qualitatively comparable to Fig.~$4$ of LB14, where similar trends of age, metallicity and $\rm [Mg/Fe]$ with $\sigma_0$ are found. The properties of satellites are shown to be indipendent of halo mass also in LB14. \par
\begin{figure}
\hspace*{-0.5cm}
\includegraphics[width=97mm]{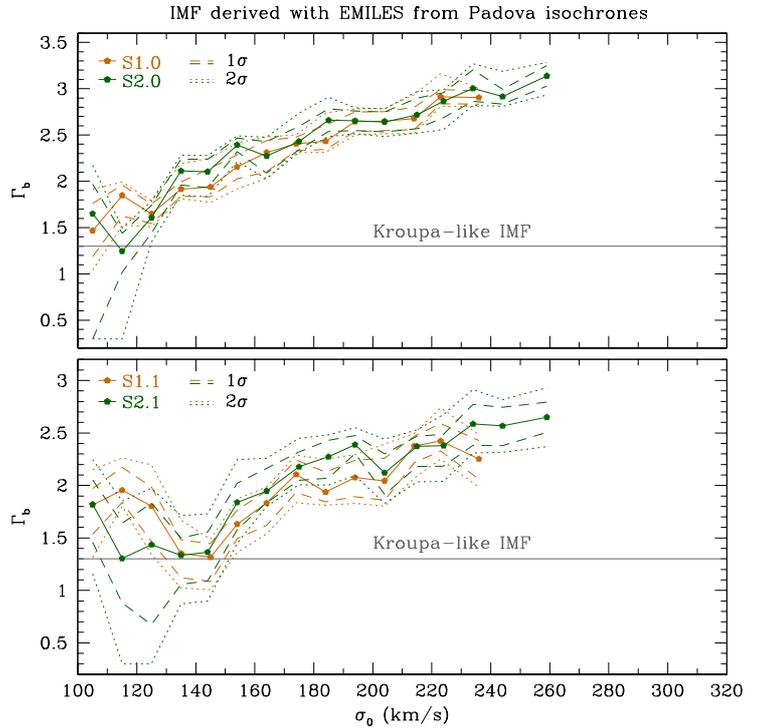}
\caption{Same as Fig.~\ref{fig:myEC12}, but for the satellite subsamples. We see no influence of the environment on the IMF slope vs. $\sigma_0$ trend, even for the lowest $\sigma_0$-bins.}
\label{fig:myES12}
\end{figure}

The trend of IMF slope with $\sigma_0$ is very similar to that for central ETGs, as shown in Fig.~\ref{fig:myES12}. The IMF becomes increasingly bottom-heavier for higher $\sigma_0$ and no trend with host halo mass is observed, within the error bars.
\subsection{Centrals and satellites}
\label{sec:cs}
\begin{figure}
\hspace*{-0.8cm}
\includegraphics[width=90mm]{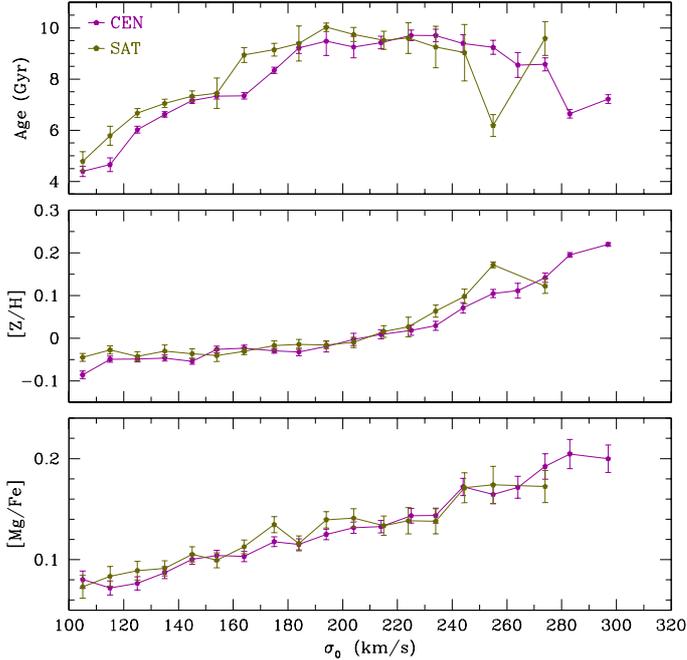}
\caption{Same as Figs.~\ref{fig:prop1} and~\ref{fig:prop3}, but comparing galaxies according to hierarchy, i.e. centrals vs. satellites.}
\label{fig:prop5}
\end{figure}

\begin{figure}
\hspace*{-0.7cm}
\includegraphics[width=97mm]{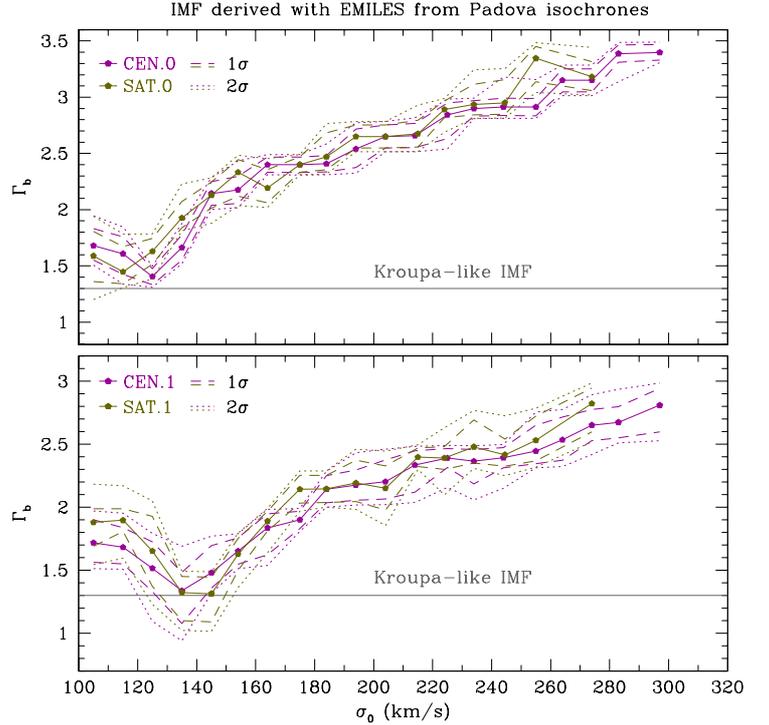}
\caption{Same as Figs.~\ref{fig:myEC12}  and~\ref{fig:myES12}, but for the CEN and SAT subsamples.}
\label{fig:myECS}
\end{figure}

Comparing centrals and satellites regardless of host halo mass yields the results shown in Fig.~\ref{fig:prop5}, for age, metallicity, and \mgfe, and Fig.~\ref{fig:myECS}, for IMF slope.\par
In general, centrals show younger ages ($\Delta Age=0.7$  Gyr), slightly more metal-poor populations ($\Delta \rm [Z/H]=0.019$ dex), and slightly lower values of $\rm [Mg/Fe]$ ($\Delta \rm [Mg/Fe]=0.01$ dex) than satellites. The difference between these two subsamples is not as pronounced as when we compare centrals in different host haloes, with the age being in fact the only property whose differences are significant at a $1.8\sigma$ level on average. Differences in metallicity and \mgfe\ can be considered not significant, since they differ only at a $\sim 1.5\sigma$ and $0.7\sigma$ level, respectively. The behaviour of age, metallicity, and \mgfe\ with central velocity dispersion mirrors that already seen for the individual subsamples of centrals and satellites.\par
Fig.~\ref{fig:myECS} shows that the IMF slope for the CEN and SAT subsamples behaves in the same way as for subsamples C1/C2 and S1/S2. At higher $\sigma_0$, the IMF becomes increasingly more bottom-heavy, while hierarchy does not affect the values of \gammab\ significantly. We also reduced the range of halo mass for the central and satellite subsamples to $\rm 10^{12.5}\le\, M_h\, < \, 10^{14}\, M_\odot/h$ to further check the robustness of our results. This test is equivalent to comparing the C2 and S1 subsamples and allows us to single out the effect of hierarchy (with respect to halo mass) on the IMF slope. We report no significant change in this comparison with respect to what shown in Fig.~\ref{fig:myECS}.
\subsection{Comparison of results from different models}
\label{sec:diffmod}
\begin{figure}
\hspace*{-1.2cm}
\includegraphics[width=97mm]{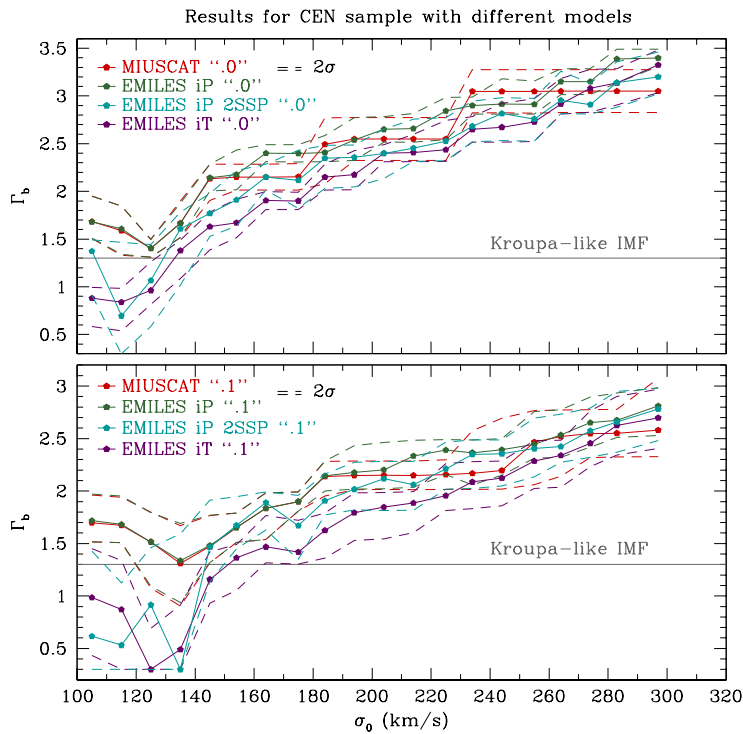}
\caption{Comparison of IMF-slope estimates obtained using different models (see labels in the top--left of each panel), for the CEN subsample. The top and bottom panels refer to the cases ``.0'' and ``.1'', respectively. ``iP'' refers to models based on Padova isochrones, while ``iT'' refers to models based on Teramo isochrones. All curves refer to 1SSP models, but the cyan lines, which have been obtained for 2SSP models (see the text). To allow a more clear comparison of different curves, the plots show only the 2$\sigma$ confindence contours on \gammab.}
\label{fig:compC}
\end{figure}
\begin{figure}
\hspace*{-0.5cm}
\includegraphics[width=97mm]{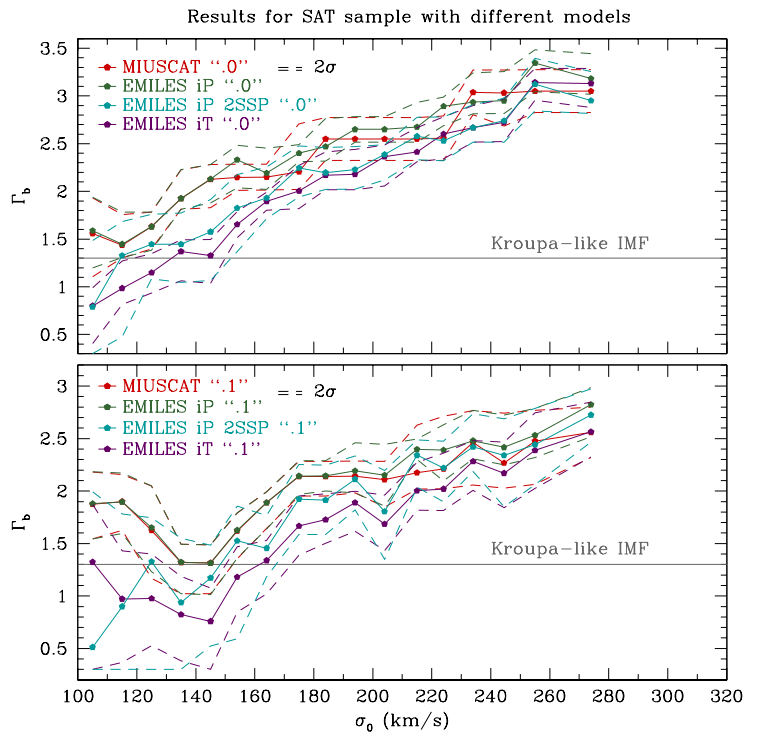}
\caption{Same as Fig.~\ref{fig:compC} but for the subsample of satellites.}
\label{fig:compS}
\end{figure}
We repeat the fitting procedure for all stacked spectra, with different sets of models. In addition to MIUSCAT and EMILES single SSP (1SSP) models constructed with Padova isochrones (hereafter EMILES iP), we also perform the fits with EMILES models based on Teramo isochrones (EMILES iT), and EMILES iP models where, instead of using a single SSP, we adopt a linear combination of two SSPs (EMILES iP 2SSP), each having different age and metallicity (treated as free fitting parameters), and the same IMF.\par
\begin{figure}
\hspace*{-1.5cm}
\includegraphics[width=105mm]{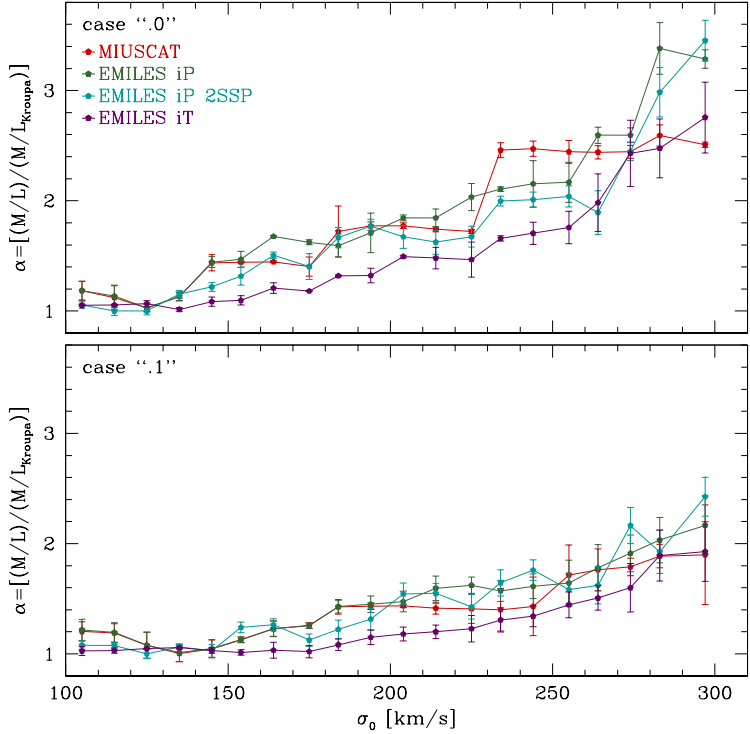}
\caption{Comparison of the mass-to-light ratio (M/L) obtained from the IMF slopes displayed in Fig.~\ref{fig:compC} and their corresponding ages and metallicities, for the CEN subsample. Each value of M/L is normalised by the M/L ratio of a stellar population with the same age and metallicity, but with a Kroupa IMF.}
\label{fig:compML}
\end{figure}

Results obtained with different models are compared in Figs.~\ref{fig:compC} and \ref{fig:compS}, for the CEN and SAT subsamples, respectively. As somewhat expected, we see minor differences between MIUSCAT and EMILES iP models. In fact, these two sets of models coincide in the optical spectral range, but for the fact that EMILES SSPs are provided for a wider range of IMF slopes, $\rm \Gamma_b$, with respect to MIUSCAT. Interestingly, both the EMILES iT and EMILES iP 2SSP models yield populations with different ages compared to EMILES iP 1SSP models:
\begin{itemize}
\item The Teramo models yield older ages with respect to the Padova ones, due to the different temperature scale of the two sets of isochrones (see \citealt{2015MNRAS.449.1177V}). This leads to lower values of \gammab\ for EMILES iT with respect to EMILES iP, especially for low $\sigma_0$ galaxies (see magenta and green curves in Figs.~\ref{fig:compC} and \ref{fig:compS}). This likely results from the degeneracy between the effects of increasing age and IMF slope on IMF sensitive indices (see e.g. LB13).
\item 2SSP models add a small fraction (see LB13) of young stars on top of a predominatly old component. This results into an IMF slope estimate which is generally lower than for 1SSP models, as seen in Figs.~\ref{fig:compC} and \ref{fig:compS} from the offset of the $\rm \Gamma_b$--$\sigma_0$ relations (cyan relative to green curves). The offset is more pronounced at low $\sigma_0$ as the fraction of frosted stars is larger (up to $\sim10$--$15 \, \%$) for low mass galaxies, while it becomes less and less important for increasing $\sigma_0$.
\end{itemize}
Finally, we note that, despite differences in the value of $\rm \Gamma_b$ at fixed $\sigma_0$, the general trend of IMF slope becoming more bottom-heavy for increasing $\sigma_0$ is well established for all models. Moreover, comparing Fig.~\ref{fig:compC} to Fig.~\ref{fig:compS} also shows that the $\rm \Gamma_b$--$\sigma_0$ relation is independent of galaxy hierarchy for all models, i.e. regardless of the adopted set of isochrones, or the assumptions on the galaxy SFH (i.e. the number of SSPs used). The same conclusion holds true also when the mass-to-light (M/L) ratios of the different samples are compared. Fig.~\ref{fig:compML} shows this comparison for our central sample CEN, where the M/L ratio shown has been normalised by a M/L ratio calculated with the same age and metallicity, but using a Kroupa IMF (this value is sometimes referred to as the mismatch parameter $\rm \alpha$ in the literature, as we do in the figure). We see that the values of M/L differ slightly from model to model and that the general trend of increasing M/L with $\rm \sigma_0$ is recovered in all models. In the calculation of the M/L ratio, a lower limit of $\rm 1.0$ has been imposed on $\rm \Gamma_b$, because the chosen set of IMF indicators is not sensitive enough to differentiate between a top-heavy and a Kroupa-like IMF. The scatter seen in the figure can be attributed to differences in the models (e.g. different isochrones), while differences in the M/L values among different methods can be attributed to the way abundances, age and metallicity are estimated from one method to the other.
\section{Discussion}
\label{sec:disc}

\subsection{Galaxy central regions and the effect of environment}
\label{sec:GalCen}
Our results are obtained for spectra observed within the SDSS fibre diameter, i.e. they refer to the galaxy central regions. In fact, at the median redshift of our galaxy sample, the SDSS fiber has a projected physical radius corresponding to $\rm \sim 0.5\, R_e$.\par
In a two-phase formation scenario (see Sec.~\ref{sec:intro}), massive ETGs are expected to form their first stars at high redshift during an intense, but short, starburst, and then accrete stars from other smaller systems via galaxy-galaxy interactions (mostly minor mergers and, on average, one major merger, \citealt{2012ApJ...744...63O,2013IAUS..295..340N}). Since minor mergers deposit the accreted material in the outskirts of a galaxy, the central regions should contain only stellar populations that have formed at high redshift during the initial starburst. Due to dynamical friction, major mergers are capable of mixing the stellar content of the two galaxies, but since the galaxies have a similar mass, we expect their central parts to have formed on average under similar physical conditions during their initial starburst at high redshift. Hence, it is reasonable to assume, in general, that the light we see in our spectra comes from stars formed in the early stages of galaxy formation.\par
Various studies observe a correlation between the central velocity dispersion, and thus the stellar mass, and the stellar population properties of ETGs (see \citealt[and references therein]{2006MNRAS.370.1106G,2006ARA&A..44..141R}). The more massive galaxies form their stars in shorter, yet more intense starbursts with respect to their less massive counterparts. More massive galaxies exhibit in this way older, metal richer stellar populations with higher values of $\rm [\alpha/Fe]$.\par
Environment plays a role in the evolution of galaxies in that it favours or disfavours star-formation (SF) depending on their hierarchy. The SF of satellites is quenched, while the SF of centrals is prolongued. We thus expect to see younger, metal-richer populations with lower $\rm [\alpha/Fe]$ in centrals with respect to satellites at fixed stellar mass \citep{2002MNRAS.337..172K,2009MNRAS.394...38P,2010MNRAS.407..937P,2011MNRAS.418L..74D,2015AN....336..505P}.\par
Finally, we expect the effect of environment to be more pronounced in more massive host haloes, since the resulting gravitational potential is stronger for these objects, the Intra-Group Medium is denser and the halo more populated.

\subsection{Comparison with LB14}
\label{sec:prop}

The trends of age, metallicity and $\rm [Mg/Fe]$ with $\sigma_0$ of our galaxy sample have been tested against evironment. These results are derived in a substantially different way from LB14. We use the line strength of specific indices to derive these stellar population properties, as opposed to the full spectral fitting performed in LB14 (using STARLIGHT; see \citealt{2005MNRAS.356..270C}). We additionally allow for the IMF slope to be a free parameter of the fit, as opposed to fixing the IMF at a Kroupa-like value ($\rm \Gamma_b=1.3$). These two different approaches produce very similar results:
\begin{itemize}
\item All subsamples have an increasing trend with central velocity dispersion for all properties, except age which flattens above $\sigma_0=200$~km/s.
\item Central galaxies show a clear trend with environment, with centrals in high mass haloes (C2) being younger, more metal-rich and having lower values of $\rm [Mg/Fe]$ with respect to centrals in low mass haloes (C1).
\item Satellite galaxies in high mass haloes (S2) show no significant difference to the limit of the given precision in their stellar population properties compared to satellites in low mass haloes (S1).
\end{itemize}
We do see however an offset of $+2$~Gyr in age, $+0.1$~dex in metallicity and $-0.05$~dex in $\rm [Mg/Fe]$, as described in Sect.\ref{sec:c12}, between our results and those of LB14. This is most likely due to the different fitting approaches used.\par
We conclude that the environmental dependence (or lack thereof) of these stellar properties of centrals (satellites) is valid against the two methods employed to analyse the same stacked spectra.\par
The trend with environment of the properties of centrals can be explained by envisaging that a more massive and thus richer host halo favours a more prolongued star formation history of its central. This would result in younger ages, more metal-rich stars and lower values of $\rm [Mg/Fe]$ in galaxies residing in more massive haloes, since more SN type Ia and type II were allowed to explode before their SF stopped, enriching the star forming gas of metals and diluting it with $\rm Fe$. The lack of trend with host halo mass of satellites can be explained if we consider the quenching of their SF to be mainly an internal process, so independent of environment, or to have occurred on a rather short time scale \citep{2005ApJ...621..673T,2010MNRAS.404.1775T,2009MNRAS.394...38P,2010MNRAS.407..937P}.\par
We note that centrals and satellites typically reside in haloes of different mass at any given $\rm \sigma_0$, thus their similar trends of stellar population properties and IMF slope as a function of $\rm \sigma_0$ should not be strongly affected by galaxy conformity (cf. for example \citealt{2006MNRAS.366....2W}).

\subsection{IMF slope}
\label{sec:slope}
The main result of the present work is the lack of dependence of the IMF slope, $\rm \Gamma_b$, on galaxy hierarchy and environment. However, we clearly detect a trend towards a more bottom-heavy IMF for ETGs with higher central velocity dispersion. This trend with $\sigma_0$ is qualitatively consistent with what was found by previous spectroscopic works \citep{2011ApJ...735L..13V,2013MNRAS.429L..15F,2013MNRAS.433.3017L,2016MNRAS.457.1468L,2017MNRAS.464.3597L,2013ApJ...776L..26C,2014MNRAS.438.1483S,2015ApJ...806L..31M,2017ApJ...841...68V,2017MNRAS.467..674T}, as well as dynamical and lensing studies of massive early-type galaxies \citep{2010ApJ...724..511A,2010ApJ...709.1195T,2012Natur.484..485C,2013ApJ...765....8T,2014ApJ...792L..37M,2015MNRAS.446..493P}. The fact that environment does not affect the IMF of ETGs suggests that this property is already established at high redshift, at least for what concerns their central regions. This is consistent with the results of \citet{2015ApJ...798L...4M} and \citet{2014ApJ...786L..10S}, who found that IMF variations at z$\sim$1 are similar to those found in nearby ETGs, based on spectroscopy and dynamics, respectively. We argue that the central parts of ETGs have their IMF set at high redshift because, in the current picture of galaxy formation and evolution, these parts form first and because only major mergers are able to mix the central content of two galaxies. Since the masses of two galaxies undergoing a major merger are similar, we expect them to have similar IMF slopes. In addition, major mergers are expected to happen only once on average in the lifetime of a galaxy, so we expect them to be incapable of considerably changing the slope of the IMF in the central parts of an ETG once that slope is set.\par
This result, especially in the case C1/C2, is in apparent tension with \cite{2015ApJ...806L..31M}, who found that there is a correlation between the metallicity and the IMF slope in the central regions of massive galaxies. The subsamples C1 and C2 have different metallicities, yet their IMF slope is similar. This apparent conflict with \cite{2015ApJ...806L..31M} can be attributed to our stellar population fitting precision and not necessarily to a real inconsistency. If we compare our results for the ``.0'' and ``.1'' cases at $\rm \sigma_0 \sim 300$~km/s, we see a difference in IMF slope of about $\rm 0.5$, which is significantly larger than the expected IMF difference in the $\rm \Gamma_b$-metallicity relation of \cite{2015ApJ...806L..31M}. In that case, an offset in metallicity of $\rm 0.05$~dex translates into an IMF slope difference of $\rm \sim 0.15$. The difference in IMF slope we observe is, however, comparable with the scatter in the relation, as seen in Fig.~2 of their paper. Thus we conclude that our results are not inconsistent with \cite{2015ApJ...806L..31M}, but rather suffer from the limitations of the current stellar population fitting precision. Furthermore, the IMF-metallicity relation presented in \cite{2015ApJ...806L..31M} refers to systems with high velocity dispersion, so that metallicity alone should not be interpreted as the fundamental driver of IMF variations.\par
Our analysis takes into account the degeneracy between changes in IMF slope and single elemental abundances. In all three comparisons (C1/C2, S1/S2, CEN/SAT), our ``.1'' fitting approach -- where abundances are fitted to the data directly -- shows in general a less steep $\sigma_0$-IMF relation, and slightly lower values of $\rm \Gamma_b$ with respect to the ``.0'' case, where only IMF sensitive features, as well as age and metallicity indicators, are fitted. Nonetheless, both methods predict an IMF more bottom-heavy than Kroupa at the highest $\sigma_0$ probed ($\rm  \sim 300$~km/s). This result is fully consistent with \citet{2015MNRAS.449L.137L}.\par
We notice here that the two subsamples of centrals divided by halo mass, C1 and C2, do not agree on the value of $\rm \Gamma_b$ for the lowest $\sigma_0$ bins ($\rm \lesssim 150$~km/s). Since this is the only case where the IMF slope is inconsistent between different subsamples, we have investigated the issue more thoroughly. After running the tests described below (Sect.~\ref{sec:probC12}), we have concluded that the discrepancy in $\rm \Gamma_b$ at low $\sigma_0$ is likely spurious, and due to a combination of three different factors: (1) some degeneracy between IMF slope and SFH; (2) the lower signal-to-noise ratio of the C2 spectra (with respect to C1); and (3) some contamination of the spectra by telluric absorption.\par
Additionally, we do not expect our definition of environment to significantly alter our results on $\rm \Gamma_b$. LB14 has shown that centrals in haloes with $\rm M_h < 10^{12.5}\, M_\odot/h$ (our C1 subsample) are mostly isolated (thus representative of very low density environments), while satellites in haloes with $M_h \ge 10^{14}\, M_\odot/h$ (S2 subsample) probe by construction high density regions. If Figs.~\ref{fig:myEC12} and~\ref{fig:myES12} are compared, no significant variation has been found when comparing C1 and S2 (see also the comparison between CEN and SAT in Fig.~\ref{fig:myECS}). This thus implies that our results likely apply also when using other environment indicators.\par
Finally, we point out that while the central regions of ETGs might indeed not be influenced by the environment where galaxies live in, as shown in the present work, their outskirts most likely are. \citet{2015MNRAS.447.1033M}, \citet{2016MNRAS.457.1468L} and \citet{2017ApJ...841...68V} have found that the IMF slope of massive elliptical galaxies shows a radial gradient (but see also \citealt{2017MNRAS.468.1594A}), varying from bottom-heavy in the centre to Kroupa-like in the outskirts, beyond a few tenths of $\rm R_e$. Therefore, it would be interesting, with the aid of ongoing integral field unit (IFU) spectroscopic surveys (e.g. MaNGa, \citealt{2015ApJ...798....7B}), to test if and how IMF radial gradients depend on environment.

\subsubsection{IMF slope of centrals at low $\sigma_0$}
\label{sec:probC12}
\begin{figure}
\hspace*{-0.5cm}
\includegraphics[width=95mm]{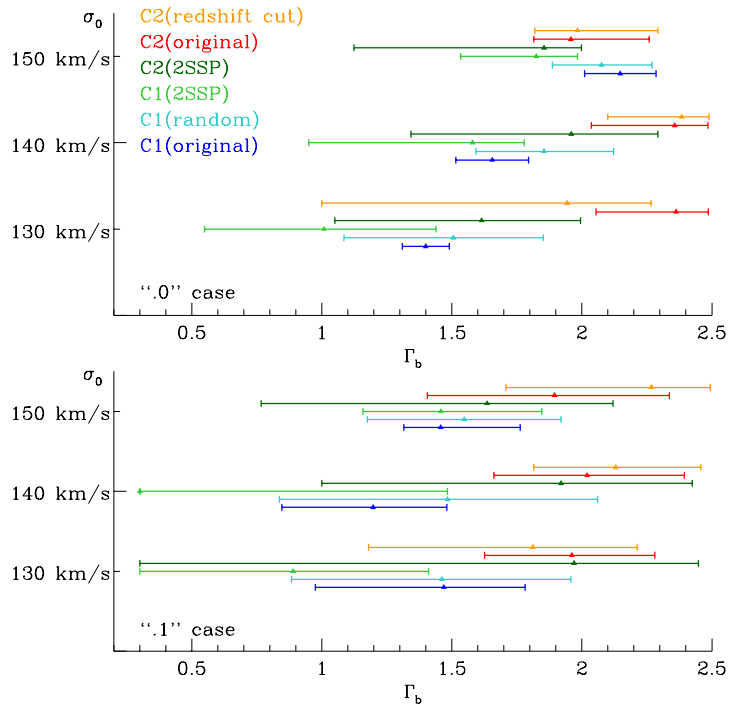}
\caption{Results of the tests performed to analyze the disagreement of IMF slope values for the three lowest-$\sigma_0$ stacked spectra of C1 and C2. The  bins, which are labeled by the upper limit of their velocity dispersion range, have a similar median velocity dispersion for both C1 and C2. We plot the values of IMF slope (points) with 2~$\sigma$ errorbars. The red and blue symbols are our original results, while the orange and cyan symbols mark, respectively, the values for which C2 has been re-stacked after excluding galaxies with redshift $z\ge 0.082$ (see text), and the values for which C1 has been randomly reduced (in 500 different iterations) to match the number of galaxies of the C2 bins. Cyan symbols give the mean IMF slope, with 2~$\sigma$ confidence intervals, from the 500 iterations. The light and dark green values are those obtained with EMILES iP 2SSP models, used to test the possible degeneracy between IMF slope and SFH. Both the ``.0'' (upper panel) and the ``.1'' case (lower panel) are shown.}
\label{fig:testC12}
\end{figure}
\begin{figure}
\hspace*{-1cm}
\includegraphics[width=97mm]{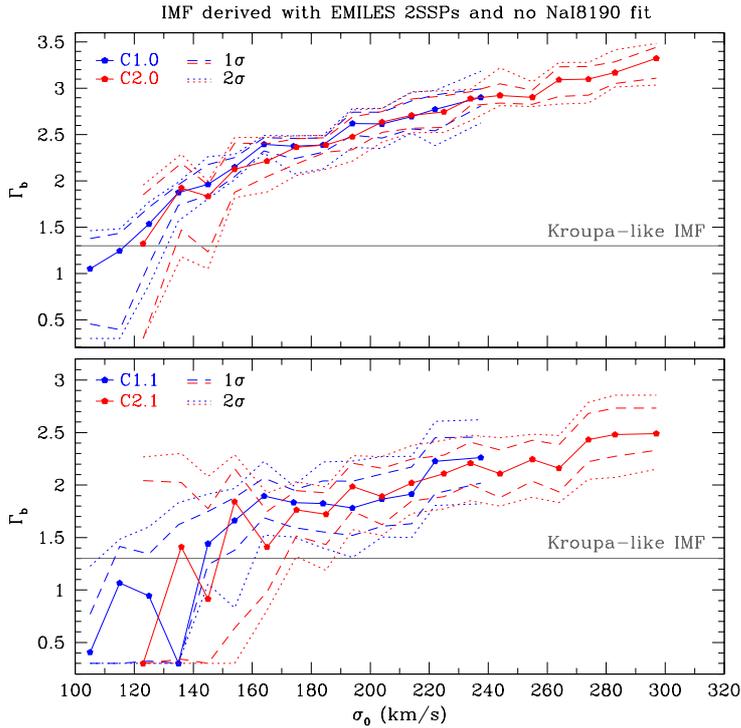}
\caption{IMF slope-$\sigma_0$ trend for C1 and C2 subsamples when 2SSP EMILES models with Padova isochrones are adopted and the $\rm NaI8190$ index is excluded from the fit. Here the difference in behaviour for C1 and C2 disappears for both fitting cases (``.0'' and ``.1'').}
\label{fig:noNa2ssp}
\end{figure}
\begin{figure*}
\includegraphics[width=130mm]{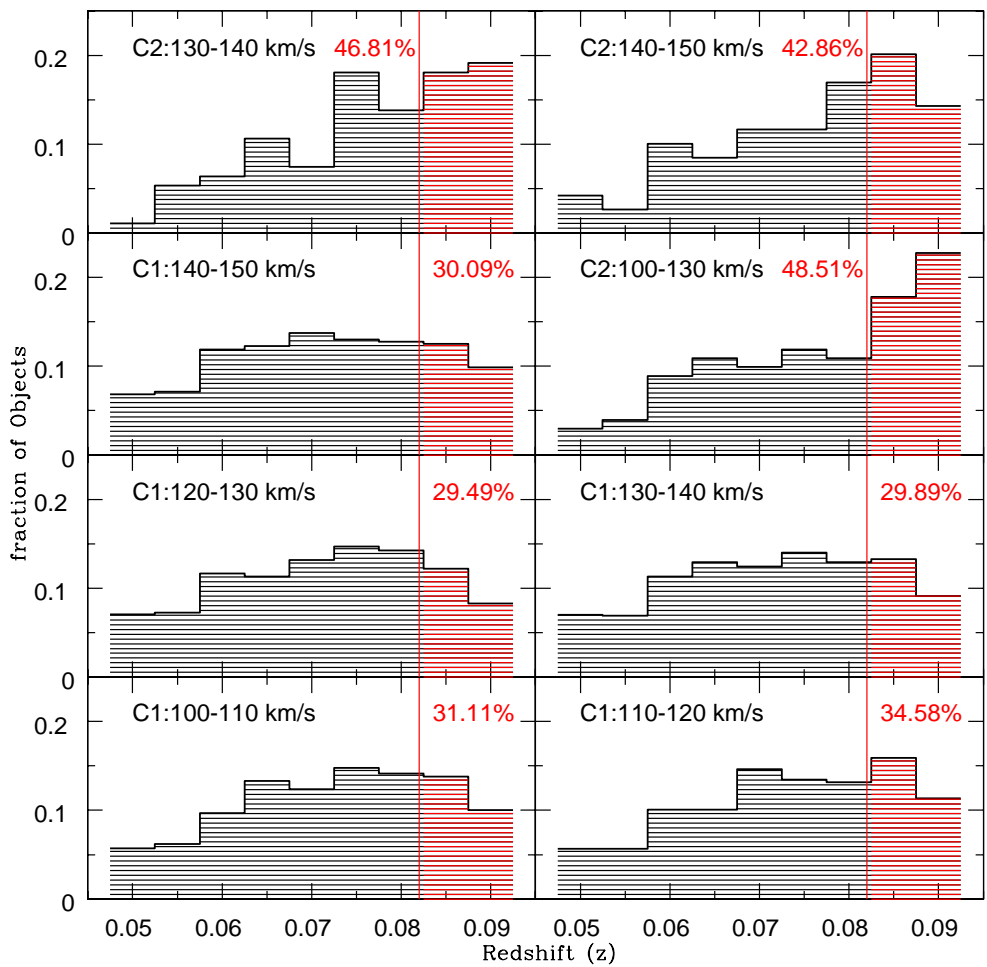}
\caption{Redshift distribution for the $\sigma_0$ bins where C1 and C2 have inconsistent values of $\rm \Gamma_b$. The vertical solid line represents the ``critical redshift'' at which telluric absorption lines enter the red pseudo-continuum of the $\rm NaI8190$ index. The red percentages refer to the number of galaxies which have a redshift higher than the critical redshift, and for which telluric absorption might be a concern.}
\label{fig:redcheck}
\end{figure*}
To investigate the origin of the discrepancy between $\rm \Gamma_b$ for the lowest $\sigma_0$  bins of subsamples C1 and C2, we focus on bins having comparable median velocity dispersion between the two subsamples and a significant discrepancy in IMF slope (i.e. $120-130$ and $130-140$ km/s for C1, $100-130$ and $130-140$ km/s for C2), as well as the first bin with good agreement between the two subsamples, taken as a reference bin ($140-150$  km/s). We performed different tests, to see if one can leave unchanged the agreement in the reference bin, while making the other $\sigma_0$ bins more consistent with each other. The results of the tests can be summarized as follows:
\begin{enumerate}
\item Assuming complex star formation histories (EMILES iP with 2SSP models) improves the agreement between the two subsamples. In fact, the sensitivity of spectral indices to variations in IMF slope decreases for lower \gammab, i.e. for lower $\sigma_0$, while their sensitivity to age, metallicity and abudance ratios remains significant (see \citealt{2013MNRAS.433.3017L}, specifically the grids in their Fig.~11). So, when taking a complex SFH into account (2SSP models) and fitting only IMF sensitive indices (``.0''  case) the discrepancy between C1 and C2 is resolved, but when we fit the abundance ratios as well (``.1''  case), the difference remains almost unchanged. Additionally, the 2 $\sigma$ contours show a high uncertainty in the values of IMF slope for the ``.1'' case (see light and dark green points in this work's Fig.~\ref{fig:testC12}). It is only because of the large uncertainties that the two subsamples ultimately agree.
\item As shown in Fig.~\ref{fig:indNaI} of Appendix~\ref{sec:ind}, the behaviour of the $\rm NaI8190$ index -- one of the main features used to constrain the IMF slope -- in the lowest $\sigma_0$-bins for the C1 and C2 subsamples is very similar to that shown by $\rm \Gamma_b$, i.e. the observed line-strengths of $\rm NaI8190$ are inconsistent between C1 and C2. In fact if we entirely remove $\rm NaI8190$ from the fit, C1 and C2 agree in even the lowest $\sigma_0$-bins in the ``.1'' case, while maintaining the same disagreement in the ``.0'' case. This is the opposite of what happens in the 2SSP test and suggests that the disagreement between C1 and C2 in the lowest $\sigma_0$ bins is caused by a combination of a more complex SFH (which is, in fact, expected for low-mass ETGs), and whatever might be affecting the strength of $\rm NaI8190$. Indeed, Fig.~\ref{fig:noNa2ssp} shows that adopting 2SSP models and excluding $\rm NaI8190$ produces fully consistent results (although with large error bars) between C1 and C2, for both fitting approaches.
\item Since the $\rm NaI8190$ index seems to be partly the reason why $\rm \Gamma_b$ does not agree between C1 and C2 at low  $\sigma_0$, we tested if contamination by telluric absorption in the spectra might be responsible for the observed discrepancy. Fig.~\ref{fig:redcheck} shows the redshift distribution in the $\sigma_0$ bins under study, marking in red the redshift range where telluric contamination can possibly affect the red pseudo-continuum of the $\rm NaI8190$ index ($z \gtrsim 0.082$). Hence, we construct new C2 stacks by excluding all galaxies at $z>0.082$. The orange points in Fig.~\ref{fig:testC12} show the results. We see an overall better agreement between C1 and C2 in the $130$ km/s bin in both ``.0'' and ``.1'', a slightly larger difference for the reference bin ($140-150$ km/s) in the ``.1'' case, and still a disagreement in the $130-140$ km/s bin for both fitting cases. Hence, telluric contamination can only explain in part the disagreement between C1 and C2.
\item Subsample C1 has a significantly larger number of objects in the four $\sigma_0$ bins from $100$ km/s to $140$  km/s ($\rm N_{C1}=4532$), resulting into a higher signal-to-noise of the stacked spectra, with respect to the two bins ($100-130$ and $130-140$ km/s) for the C2 subsample ($\rm N_{C2}=195$). We test the impact of signal-to-noise ratio and subsample size on the estimation of IMF slope by randomly reducing the C1 subsample so as to match both the number of  galaxies and the $\sigma_0$ range of the C2 bins. This ``random'' resampling is performed 500 times to obtain a significant statistics of possible outcomes. As shown in Fig.~\ref{fig:testC12}, the results of this test (cyan points) imply an overall better agreement with the original C2 values (red points) in the  ``.1'' case. For the ``.0'' case, the $130$ km/s bin shows only a slightly better agreement, while the other two bins are fully consistent, with C2. It is interesting to notice that in contrast to what happens for (iii), it is the $130$ km/s bin that does not change much from its original value, while the $130-140$ km/s bin shifts to values consistent with C2.
\end{enumerate}
In summary, we conclude that the reason why C1 and C2 do not agree in their values of IMF slope in the lowest $\sigma_0$ bins is likely spurious, and results from a complex combination of different effects. To definitively solve the issue, we would need a larger subsample of C2 galaxies with a redshift distribution similar to that of C1.

\section{Conclusions}
\label{sec:conc}
We select our galaxy sample by cross-matching the SPIDER bona-fide ETGs catalogue of \citet{2014MNRAS.445.1977L} to the group catalogue of \citet{2014MNRAS.439..611W}, which provides galaxy hierarchy and the mass of the dark matter host halo assigned to the group the galaxy belongs to. Our final sample consists of 20996 ETGs with SDSS 1d-spectra available. We stack the spectra in bins of central velocity dispersion after dividing them into subsamples based on their hierarchy and host halo mass. We measure a set of absorption line indices to constrain age, metallicity, $\rm [Mg/Fe]$, and IMF slope, by fitting the equivalent widths of these indices with predictions from state-of-the-art synthetic stellar population models (EMILES).\par
Our results are presented in Figs.~\ref{fig:prop1}-\ref{fig:myECS} and can be summarized as follows:
\begin{itemize}
\item Age,$\rm [Z/H]$ and $\rm [Mg/Fe]$: The general trend of all subsamples shows an increase of these properties with central velocity dispersion, with the exception of age above $\sigma=200$~km/s, which shows a flat dependence. Centrals in higher mass haloes are typically younger, more metal-rich and having lower values of $\rm [Mg/Fe]$ with respect to centrals residing in lower mass haloes. Satellite galaxies in different host haloes are, on the other hand, not influenced by environment to the limit of our precision: their age, metallicity and $\rm [Mg/Fe]$ are on average compatible within their respective errors. Independently of host halo mass, centrals appear on average younger than satellites. Differences in metallicity between the two subsamples are borderline significant, while their values of $\rm [Mg/Fe]$ are compatible within their respective errors.
\item IMF slope: For all subsamples, the IMF slope follows a clear trend with central velocity dispersion $\sigma_0$; the higher $\sigma_0$, the more bottom-heavy the IMF. This trend is robust against a number of ingredients (e.g. the set of isochrones used to construct the stellar population models). Hierarchy and host halo mass do not affect significantly the IMF slope. The disagreement between C1 and C2 at low $\sigma_0$ is likely spurious, resulting from a complex combination of different factors (a more complex SFH  at low $\sigma_0$, lower signal-to-noise of the C2 subsample, some contamination by telluric lines in the stacked spectra).
\end{itemize}
We conclude that, while effects of galaxy environment can be observed in the average behaviour of age, metallicity, and $\rm [Mg/Fe]$, the shaping of the IMF slope in the central parts of ETGs is settled in the early stages of galaxy formation, by processes that are not significantly influenced by environment or hierarchy at present day, such as the complex modes of star formation in the central regions of these systems in the early stages of their formation. Our motivation to separate external from internal processes (i.e. processes within host halo and galaxy scale, respectively) should be viewed as a first order approach to understand the role of the different drivers of IMF variations.

\section*{Acknowledgements}
GR and AP warmly thank Eva Grebel for logistic and financial support during this project. FLB acknowledges financial support from the visitor programm of the Sonderforschungsbereich SFB 881 ``The Milky Way System'' of the German Research Foundation (DFG). FLB acknowledges support from grant AYA2016-77237-C3-1-P from the Spanish Ministry of Economy and Competitiveness (MINECO). The authors would like to thank Scott Trager for useful comments.




\bibliographystyle{mnras}
\bibliography{biblio} 


\clearpage
\appendix

\section{Trend of measured indices with $\sigma_0$}
\label{sec:ind}
\begin{figure}
\hspace*{-0.8cm}
\includegraphics[width=95mm]{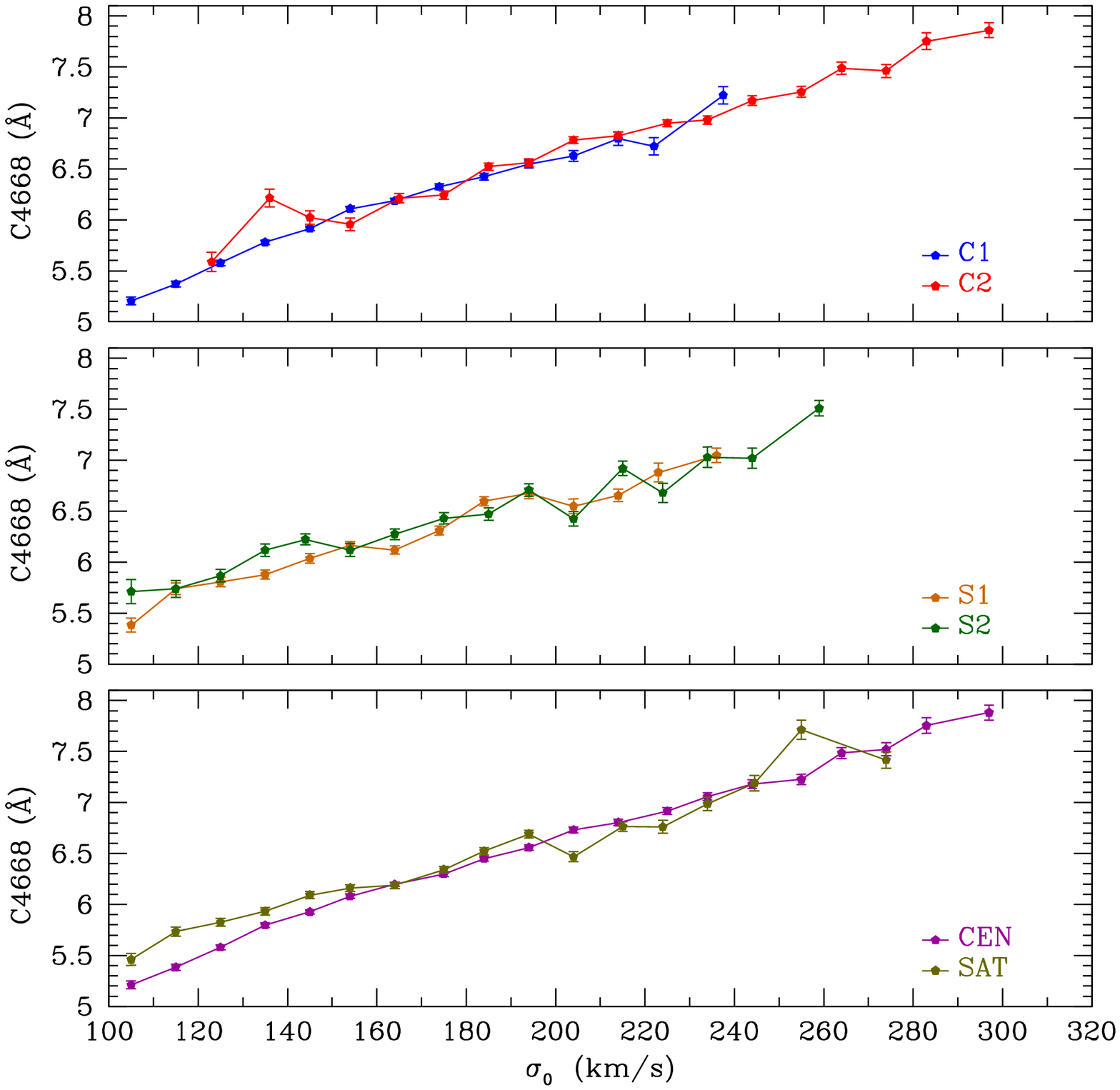}
\caption{Trend of the $\rm C4668$ line with $\sigma_0$ for the stacked spectra of C1/C2, S1/S2 and CEN/SAT, respectively. $\rm C4668$ is mainly used to fit $\rm [C/Fe]$ in the ``.1'' case.}
\label{fig:indC}
\end{figure}

\begin{figure}
\hspace*{-0.8cm}
\includegraphics[width=95mm]{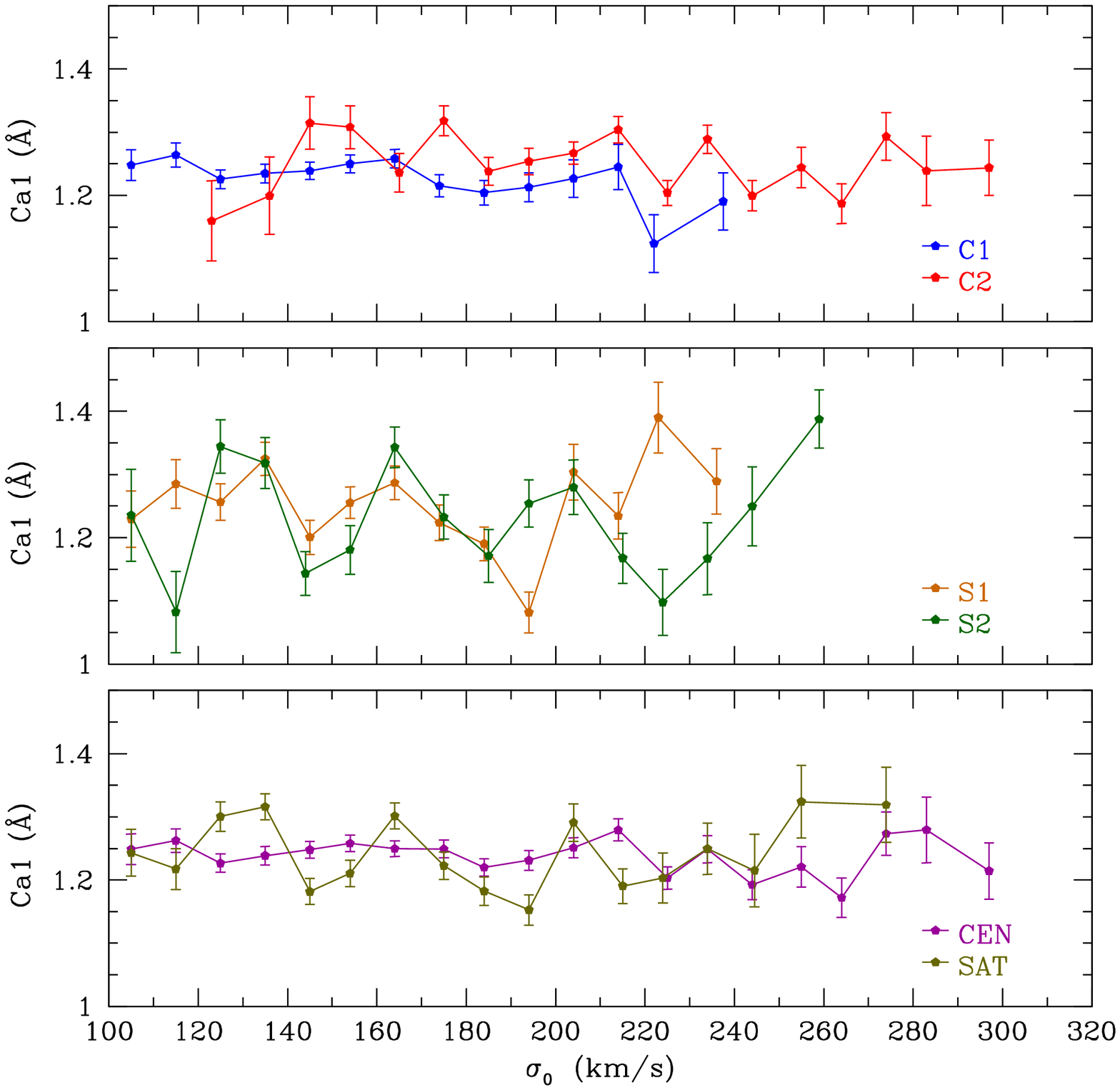}
\caption{Trend of the $\rm Ca1$ line with $\sigma_0$. $\rm Ca1$ is mainly used to fit $\rm [Ca/Fe]$, jointly with $\rm Ca4227$ and $\rm CaH\&K$, in the ``.1'' case.}
\label{fig:indCa1}
\end{figure}

\begin{figure}
\hspace*{-1cm}
\includegraphics[width=95mm]{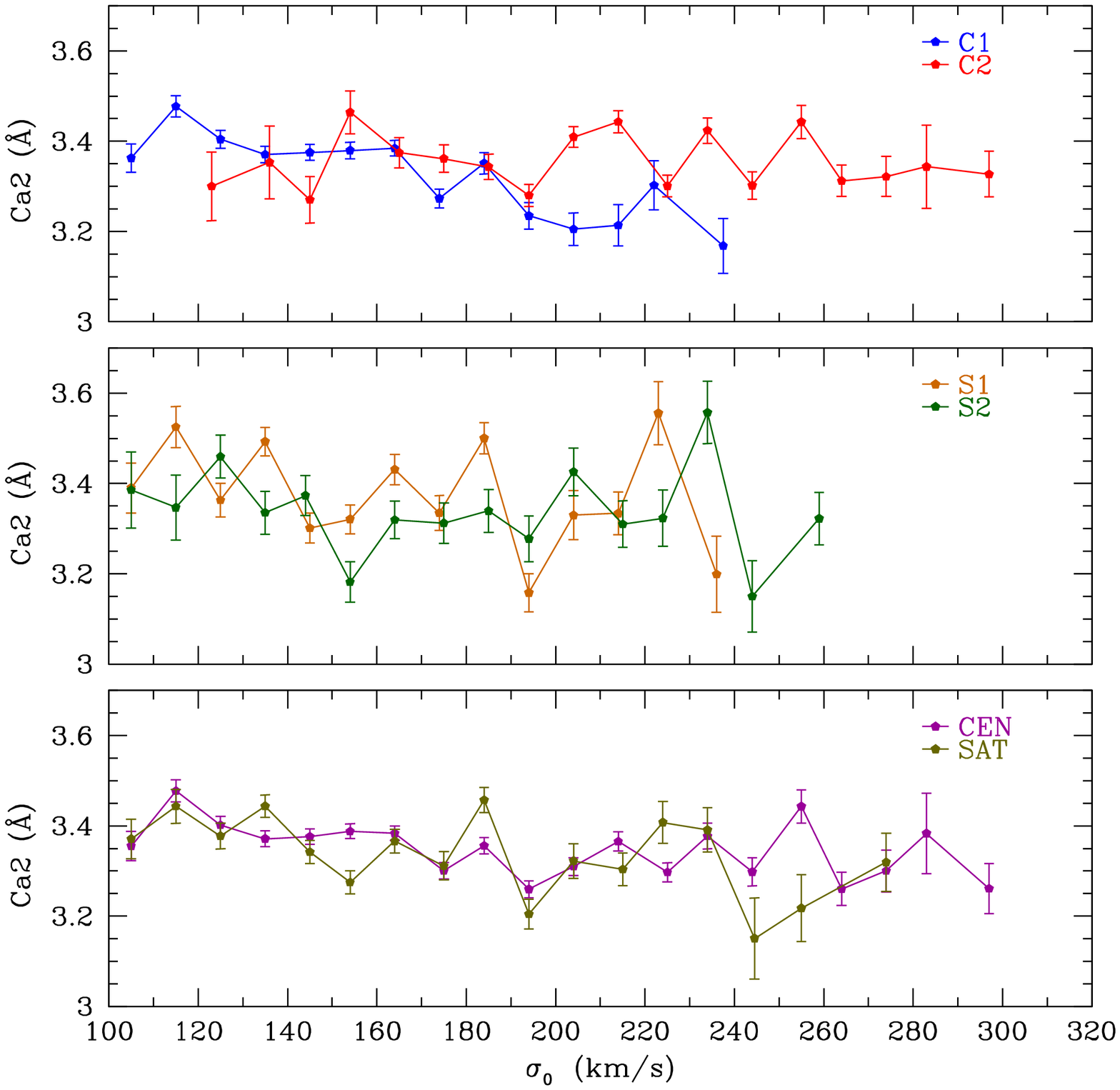}
\caption{Trend of the $\rm Ca2$ line with $\sigma_0$. $\rm Ca2$ is mainly used to fit the slope of the IMF.}
\label{fig:indCa2}
\end{figure}

\begin{figure}
\hspace*{-1cm}
\includegraphics[width=95mm]{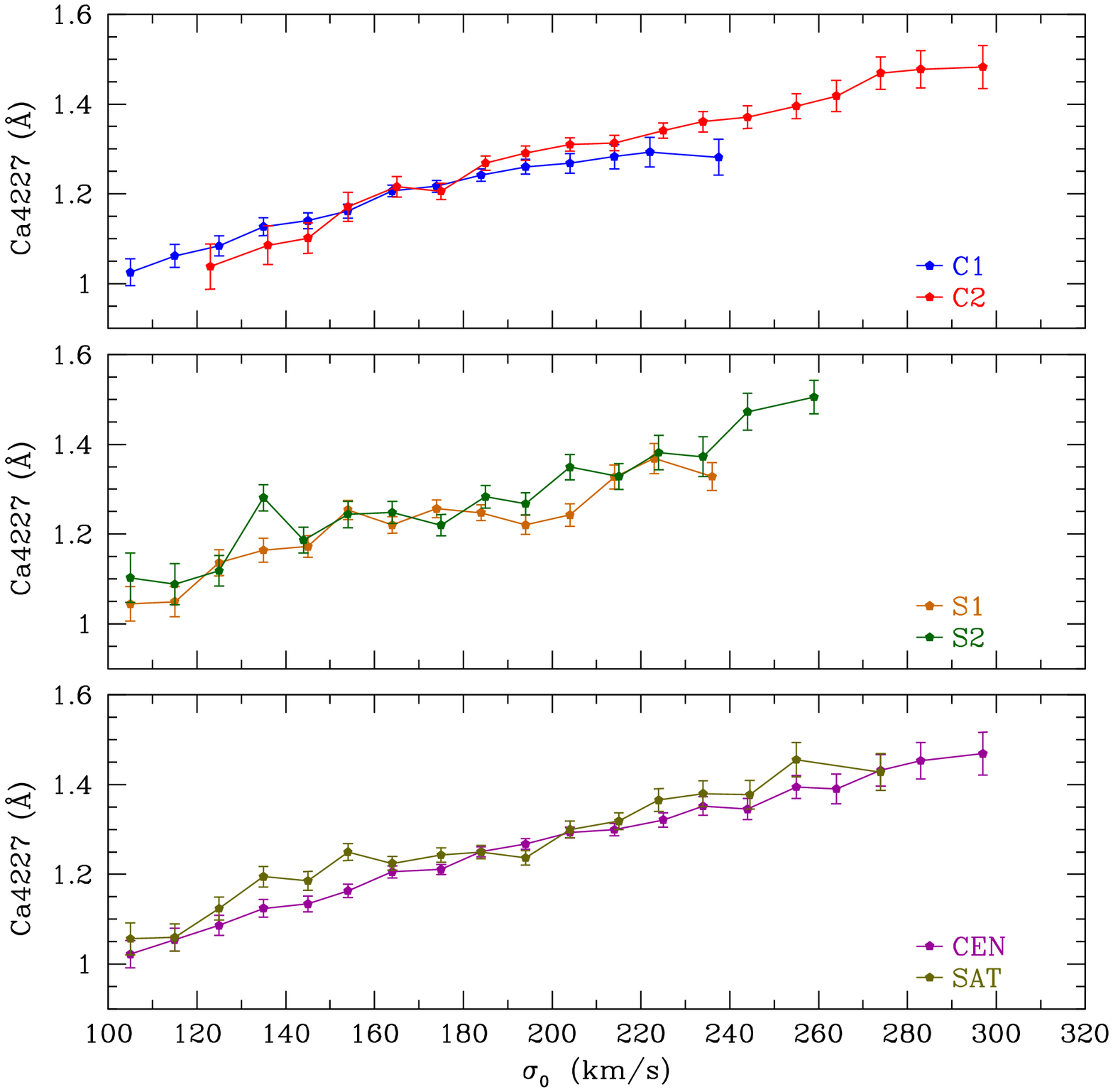}
\caption{Trend of the $\rm Ca4227$ line with $\sigma_0$. $\rm Ca4227$ is mainly used to fit $\rm [Ca/Fe]$, jointly with $\rm Ca1$ and $\rm CaH\&K$, in the ``.1'' case.}
\label{fig:indCa}
\end{figure}

\begin{figure}
\hspace*{-0.8cm}
\includegraphics[width=95mm]{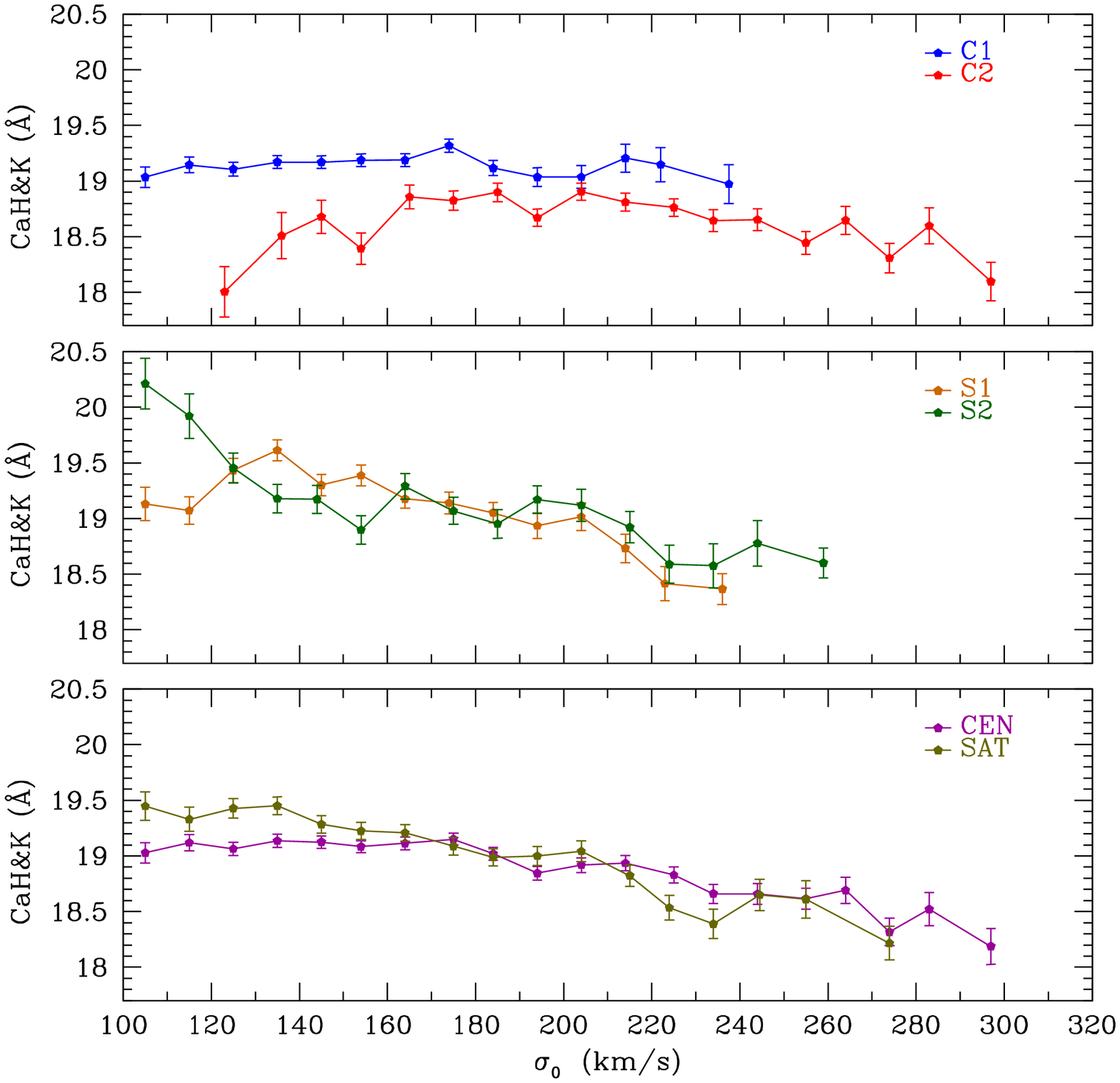}
\caption{Trend of the $\rm CaH\&K$ line with $\sigma_0$. $\rm CaH\&K$ is mainly used to fit the slope of the IMF and $\rm [Ca/Fe]$, jointly with $\rm Ca1$ and $\rm Ca4227$.}
\label{fig:indCaII}
\end{figure}

\begin{figure}
\hspace*{-0.8cm}
\includegraphics[width=95mm]{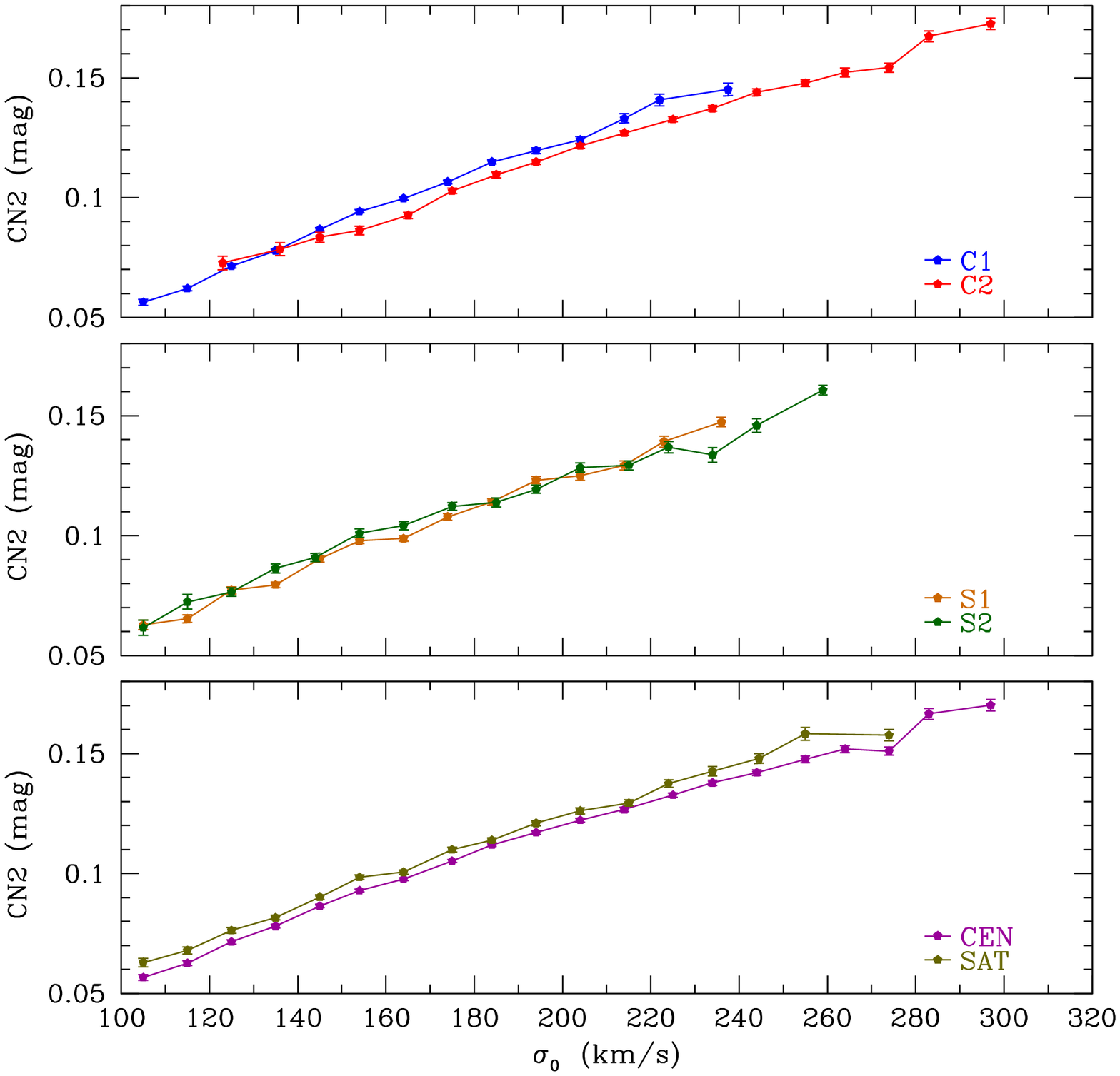}
\caption{Trend of the $\rm CN2$ line with $\sigma_0$. $\rm CN2$ is mainly used to fit $\rm [N/Fe]$, $\rm [O/Fe]$ and, jointly with $\rm C4668$, $\rm [C/Fe]$ in the ``.1'' case.}
\label{fig:indCN2}
\end{figure}

\begin{figure}
\hspace*{-1cm}
\includegraphics[width=95mm]{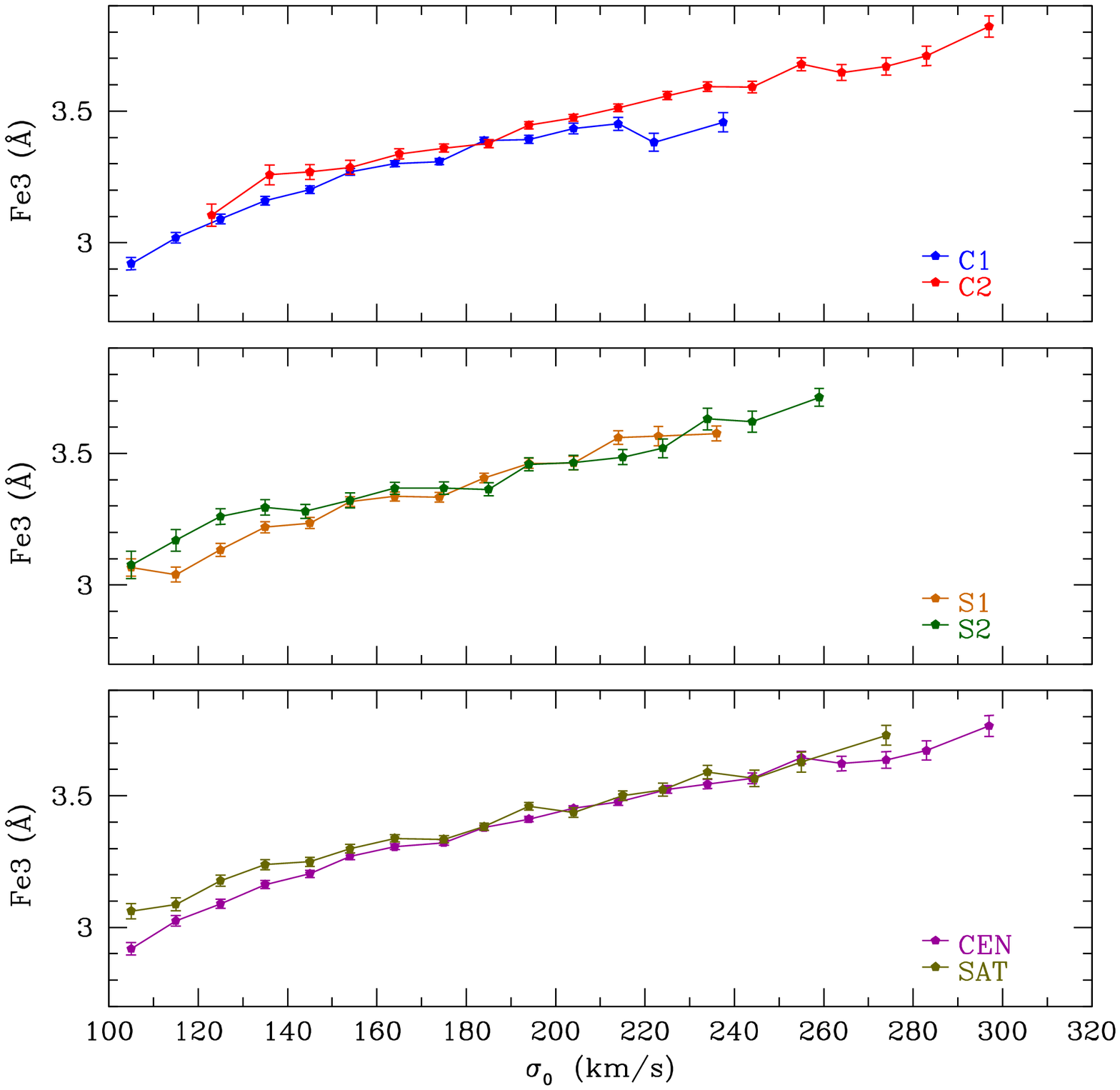}
\caption{Trend of the $\rm Fe3$ line with $\sigma_0$. $\rm Fe3$ is mainly used, jointly with $\rm Mgb5177$, to determine the value of the $\rm [\alpha/Fe]$ proxy of the stacked spectrum.}
\label{fig:indFe3}
\end{figure}

\begin{figure}
\hspace*{-1cm}
\includegraphics[width=95mm]{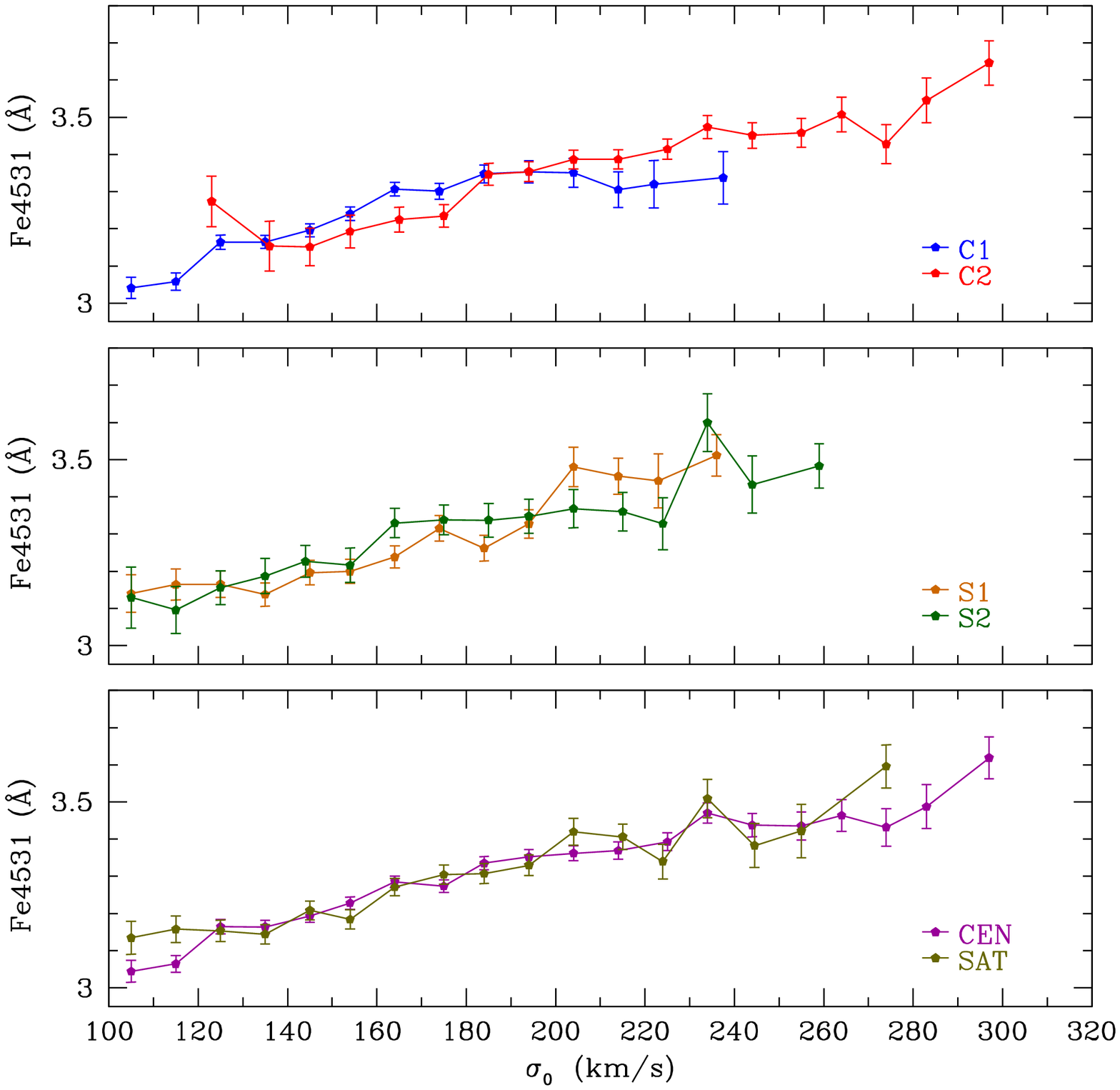}
\caption{Trend of the $\rm Fe4531$ index with $\sigma_0$. $\rm Fe4531$ is mainly used to fit $\rm [Ti/Fe]$, jointly with $\rm TiO1$ and $\rm TiO2_{SDSS}$, in the ``.1'' case.}
\label{fig:indFe}
\end{figure}

\begin{figure}
\hspace{-0.8cm}
\includegraphics[width=95mm]{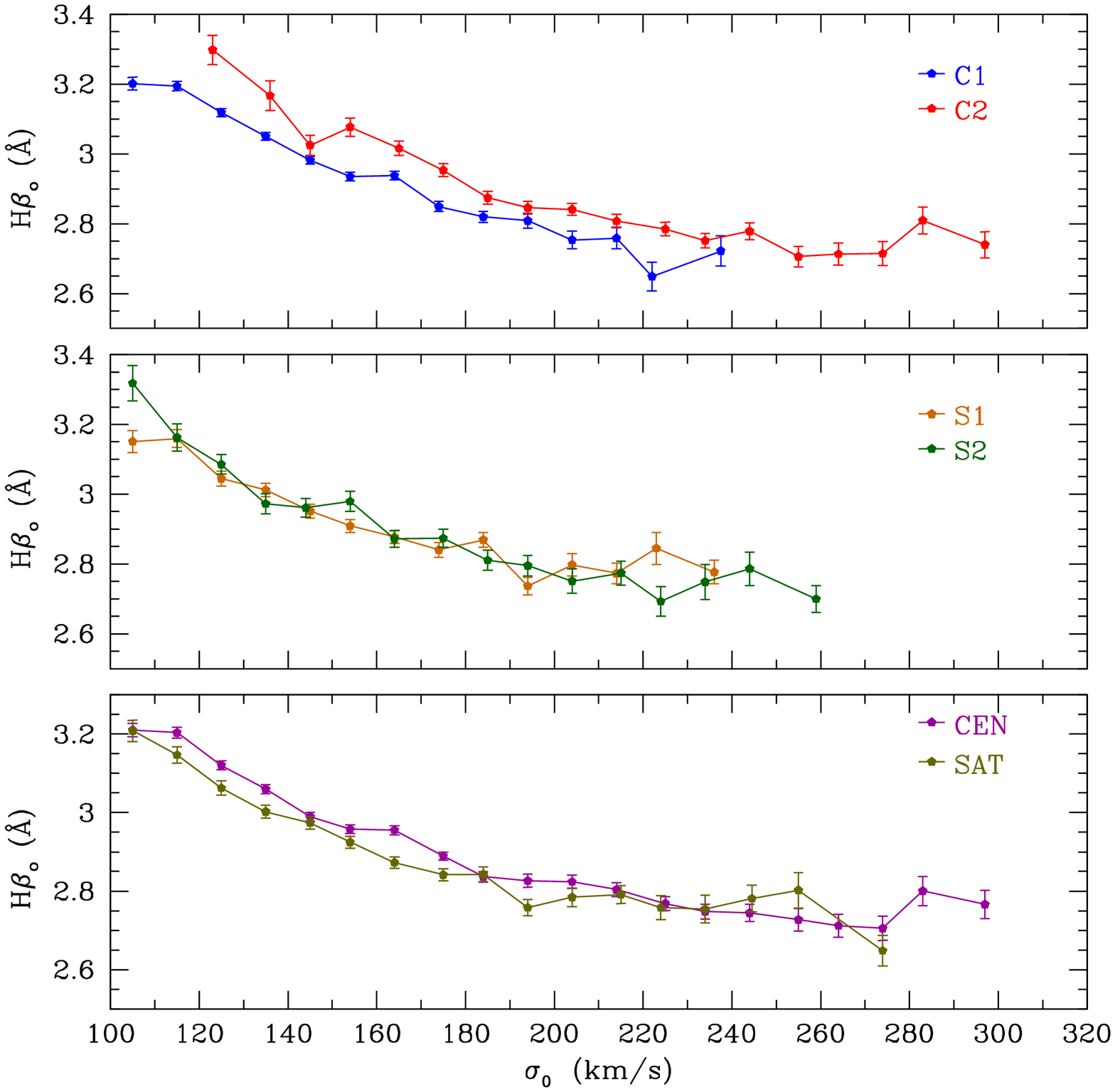}
\caption{Trend of the $\rm H\beta_o$ line with velocity dispersion $\sigma_0$. $\rm H\beta_o$ is mainly used as a proxy for the age of the stellar population of the stacked spectrum. The values of $\rm H\beta_o$ shown are corrected for nebular emission.}
\label{fig:indHbo}
\end{figure}

\begin{figure}
\hspace*{-0.8cm}
\includegraphics[width=95mm]{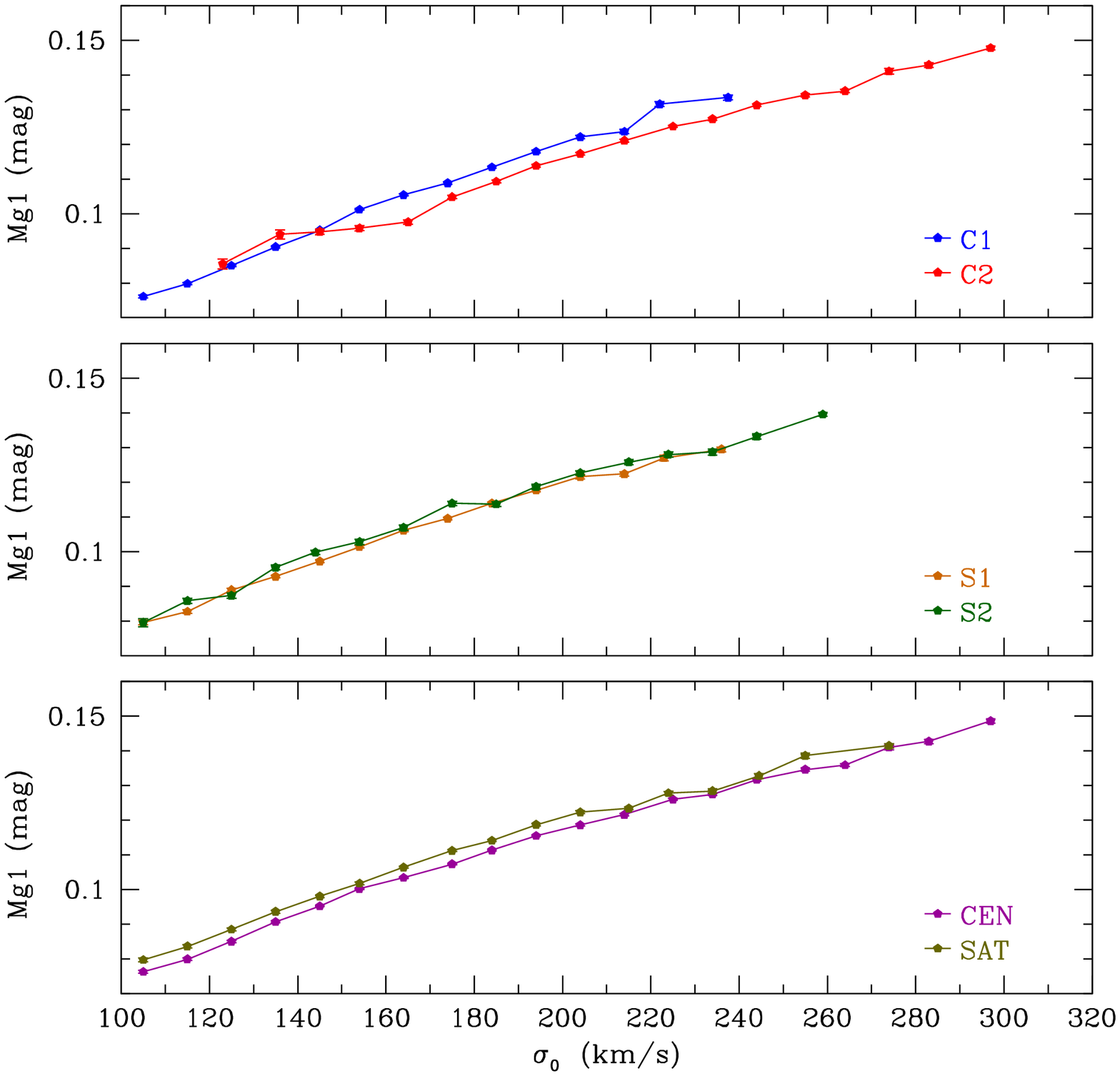}
\caption{Trend of the $\rm Mg1$ line with $\sigma_0$. $\rm Mg1$ is mainly used to fit $\rm [O,Ne,S/Fe]$, $\rm [C/Fe]$ and $\rm Si/Fe$ in the ``.1'' case.}
\label{fig:indMg1}
\end{figure}

\begin{figure}
\hspace*{-1cm}
\includegraphics[width=95mm]{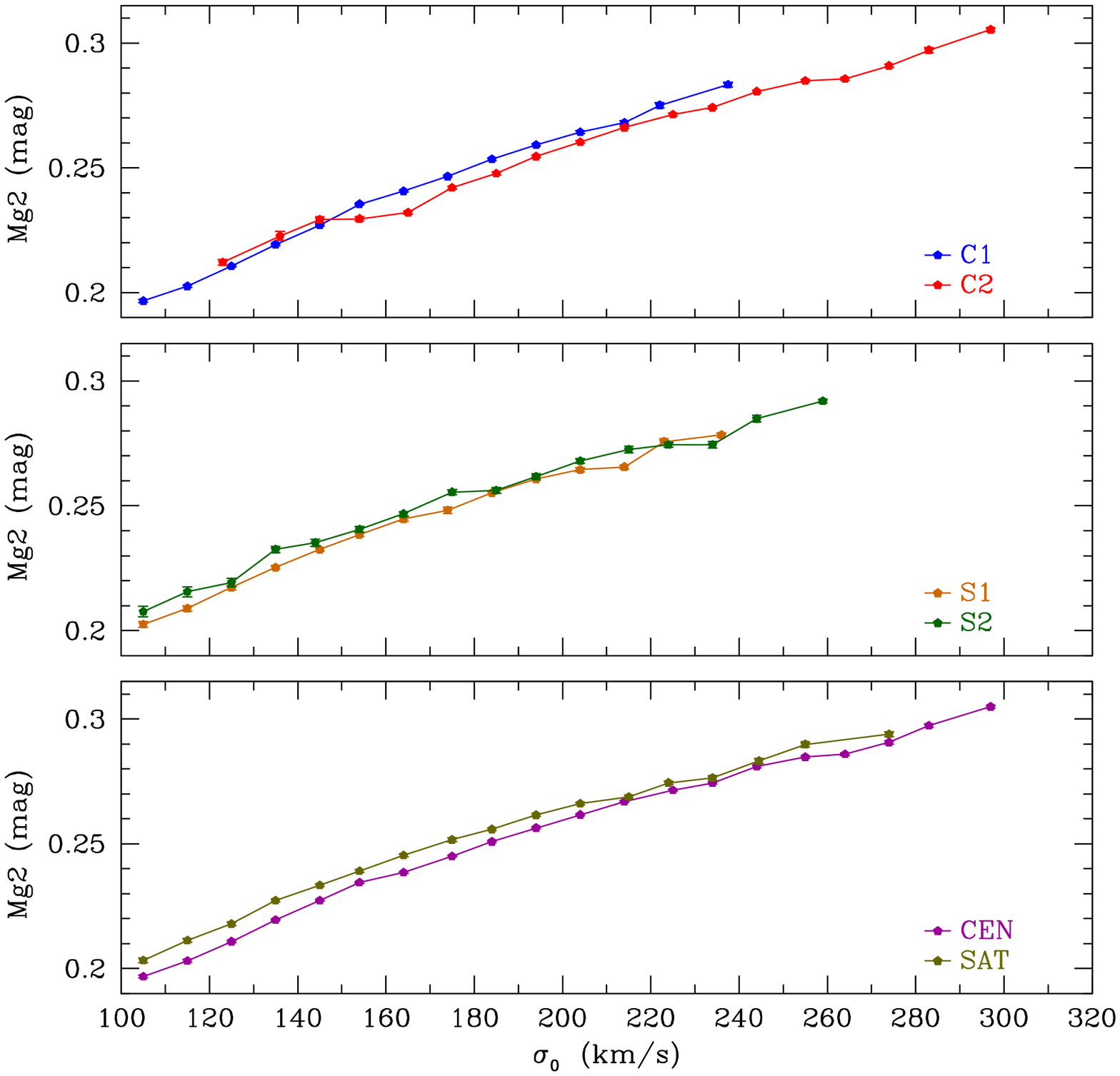}
\caption{Trend of the $\rm Mg2$ line with $\sigma_0$. $\rm Mg2$ is mainly used to fit $\rm [Mg/Fe]$ and $\rm [Si/Fe]$ in the ``.1'' case.}
\label{fig:indMg2}
\end{figure}

\begin{figure}
\hspace*{-1cm}
\includegraphics[width=95mm]{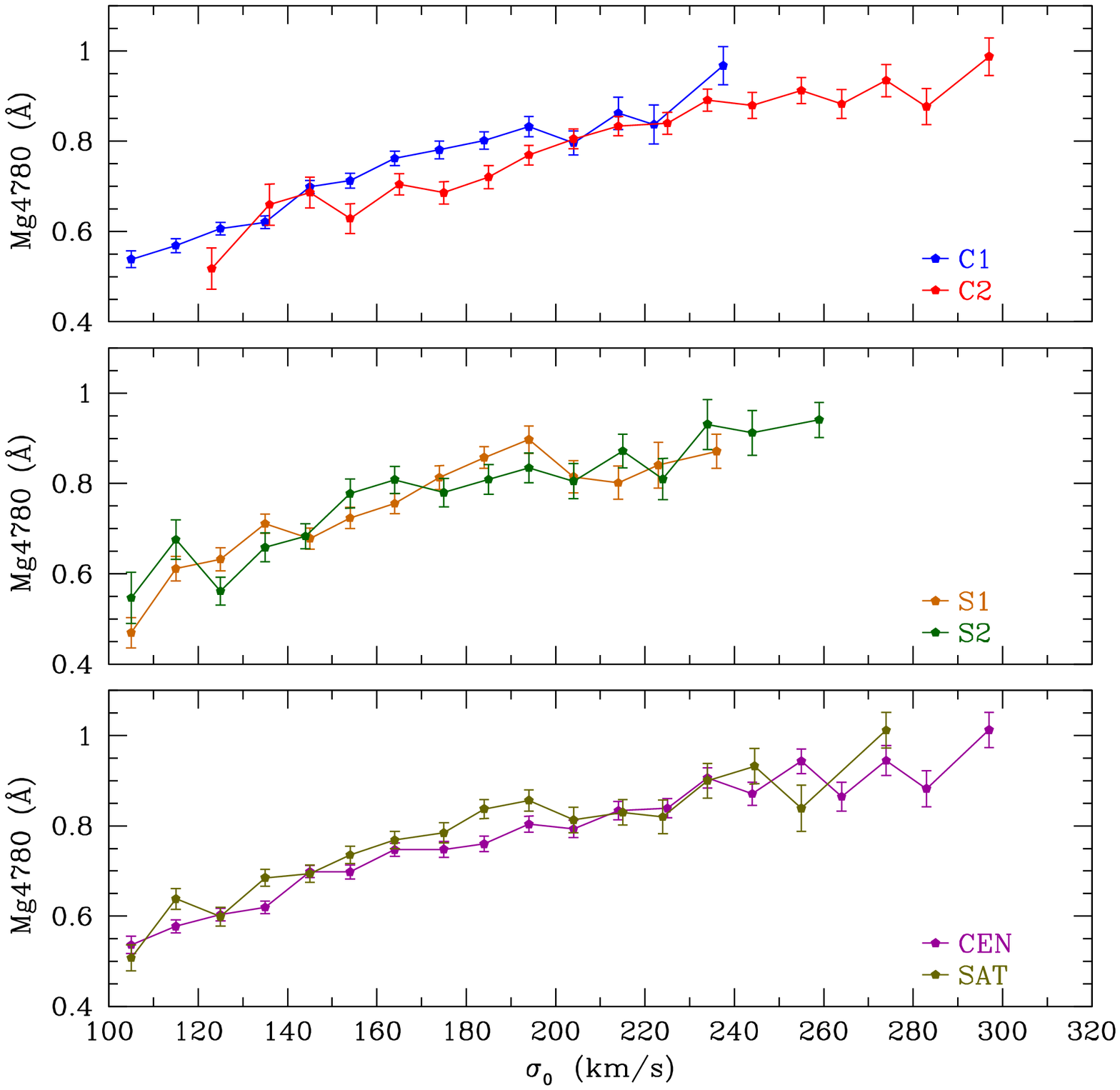}
\caption{Trend of the $\rm Mg4780$ line with $\sigma_0$. $\rm Mg4780$ is mainly used to fit the slope of the IMF.}
\label{fig:indMg}
\end{figure}

\begin{figure}
\hspace*{-0.8cm}
\includegraphics[width=95mm]{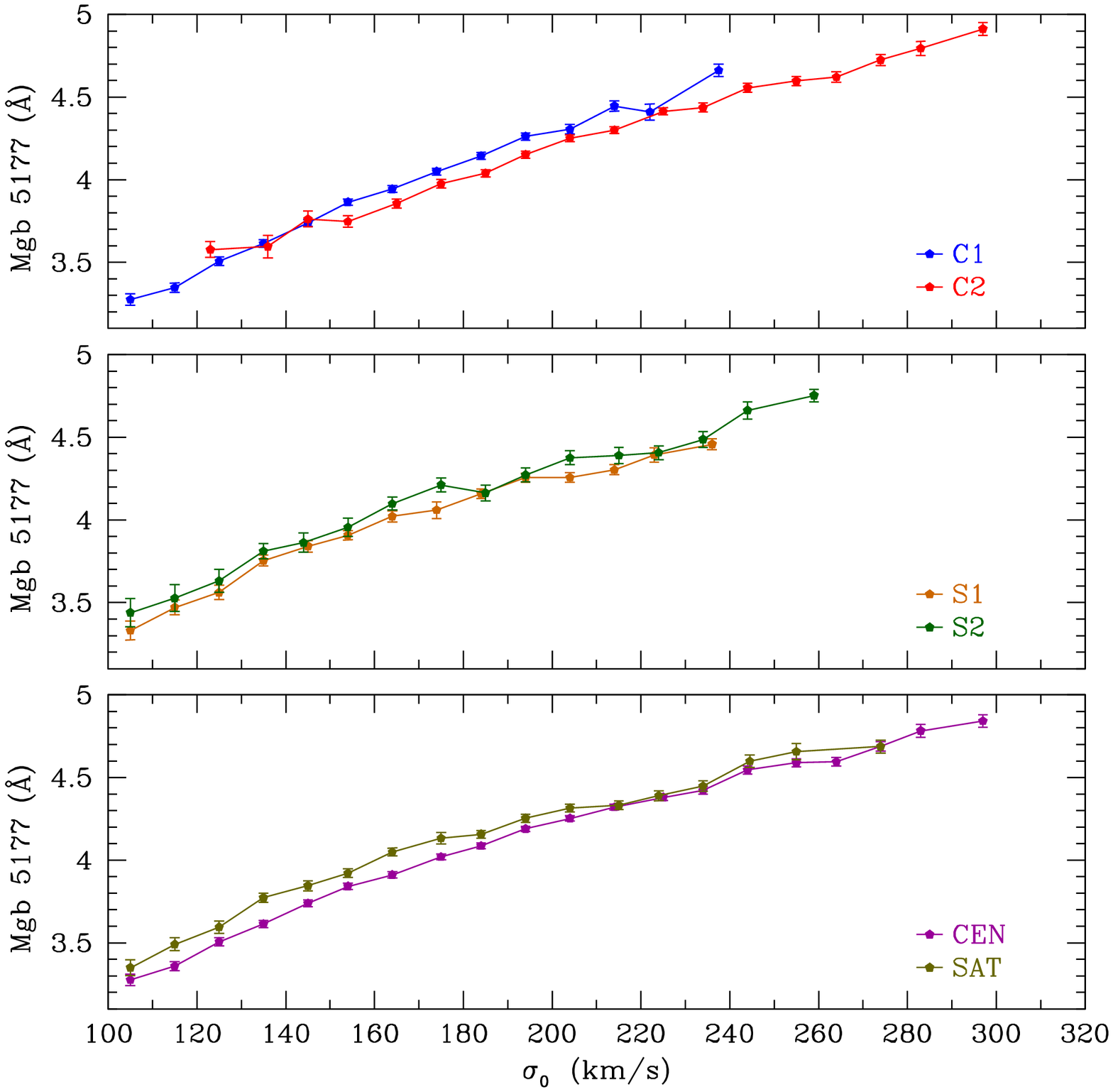}
\caption{Trend of the $\rm Mgb5177$ line with $\sigma_0$. $\rm Mgb5177$ is mainly used to compute $\rm [MgFe]'$, to fit $\rm [Mg/Fe]$ and, jointly with $\rm Fe3$, to determine the proxy of $\rm [\alpha/Fe]$.}
\label{fig:indMgb}
\end{figure}

\begin{figure}
\hspace*{-0.8cm}
\includegraphics[width=95mm]{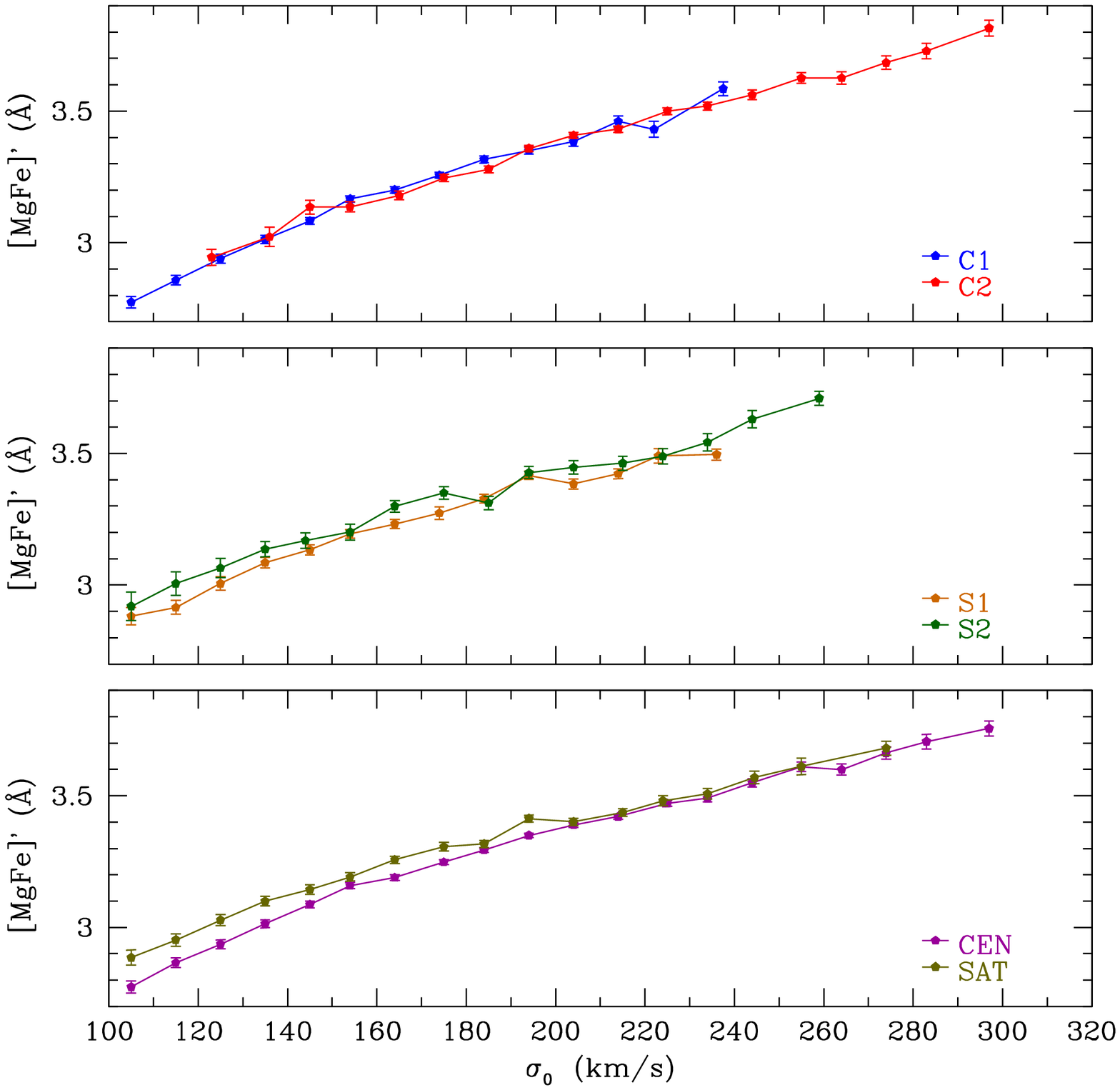}
\caption{Trend of the $\rm [MgFe]'$ line with $\sigma_0$. $\rm [MgFe]'$ is mainly used as a proxy for the metallicity of the stacked spectrum.}
\label{fig:indMgFep}
\end{figure}

\begin{figure}
\hspace*{-1cm}
\includegraphics[width=95mm]{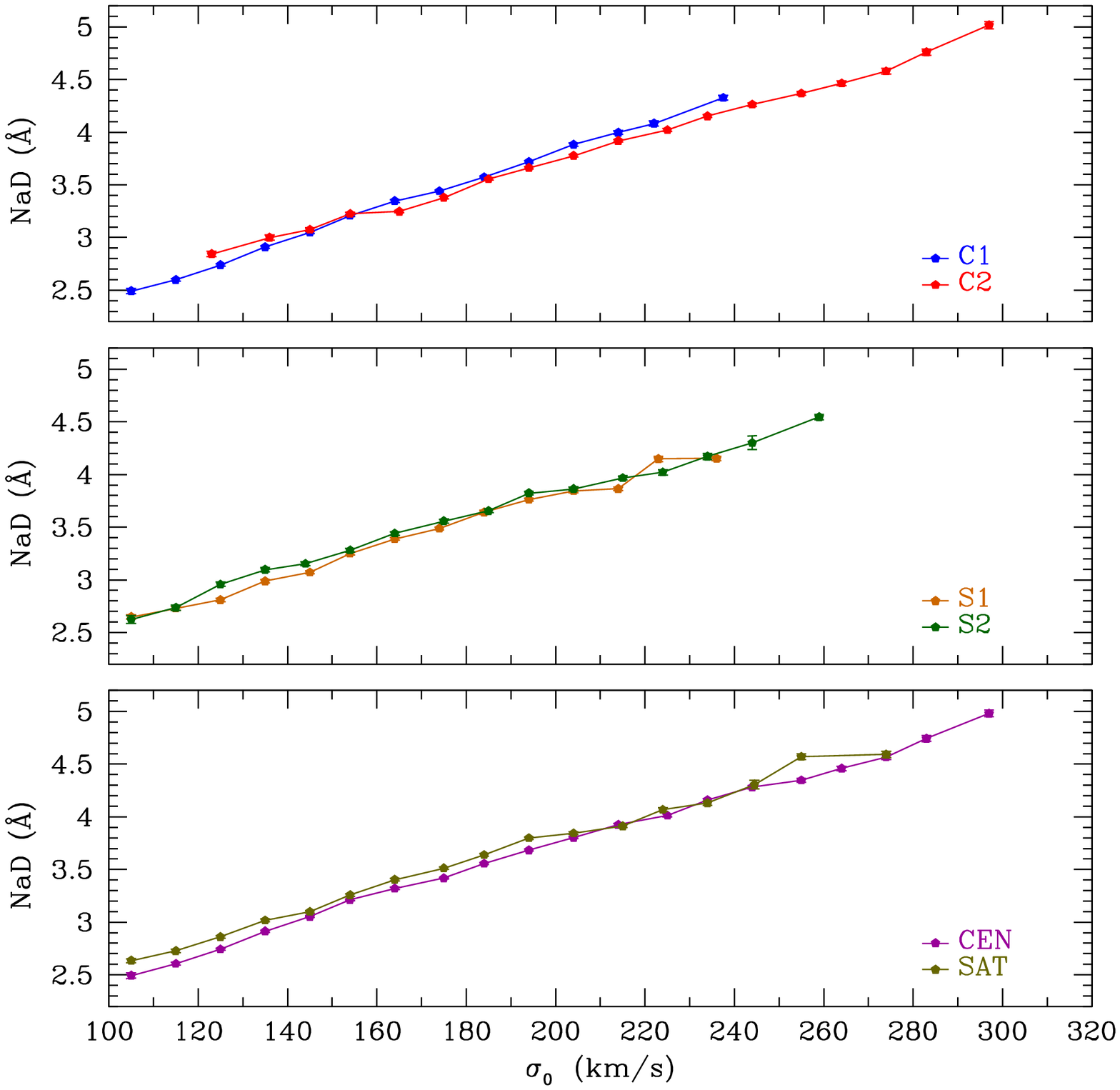}
\caption{Trend of the $\rm NaD$ line with $\sigma_0$. $\rm NaD$ is mainly used to fit the slope of the IMF and to fit $\rm [Na/Fe]$, jointly with $\rm NaI8190$.}
\label{fig:indNaD}
\end{figure}

\begin{figure}
\hspace*{-1cm}
\includegraphics[width=95mm]{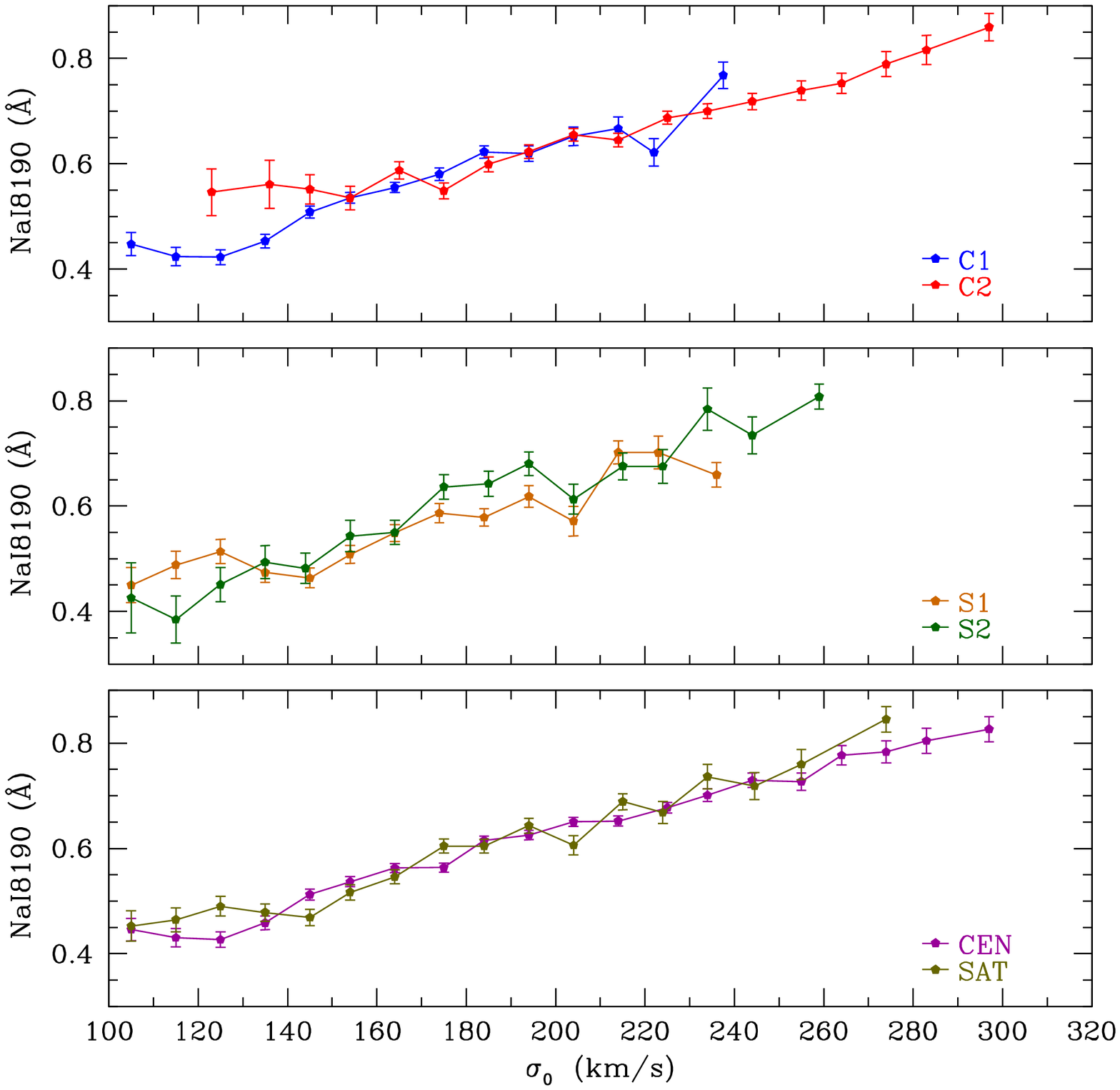}
\caption{Trend of the $\rm NaI8190$ line with $\sigma_0$. $\rm NaI8190$ is mainly used to fit the slope of the IMF and $\rm [Na/Fe]$, jointly with $\rm NaD$.}
\label{fig:indNaI}
\end{figure}

\begin{figure}
\hspace*{-0.8cm}
\includegraphics[width=95mm]{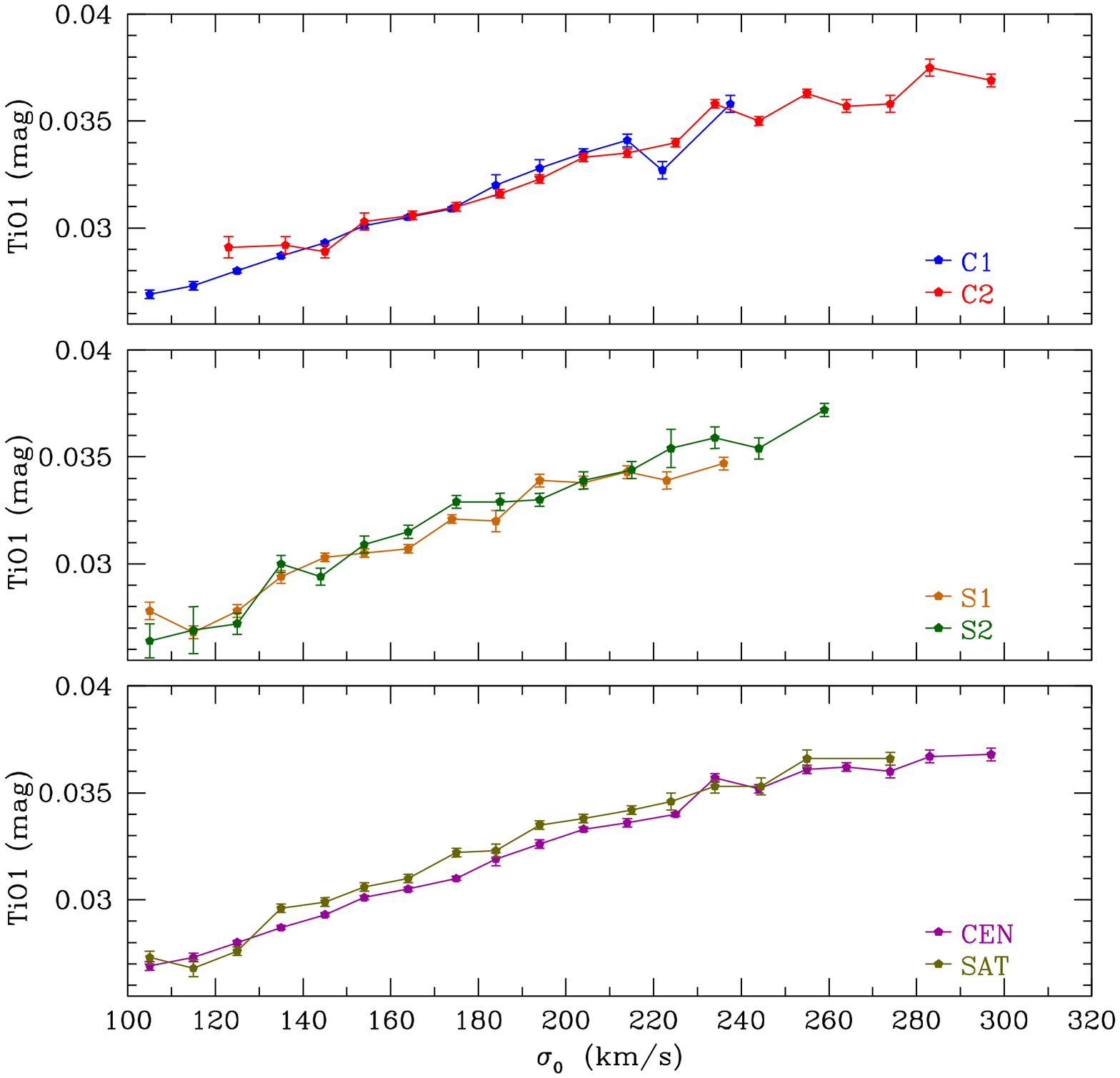}
\caption{Trend of the $\rm TiO1$ line with $\sigma_0$. $\rm TiO1$ is mainly used to fit the slope of the IMF and $\rm [Ti/Fe]$, jointly with $\rm TiO2_{SDSS}$ and $\rm Fe4531$.}
\label{fig:indTiO1}
\end{figure}

\begin{figure}
\hspace*{-0.8cm}
\includegraphics[width=95mm]{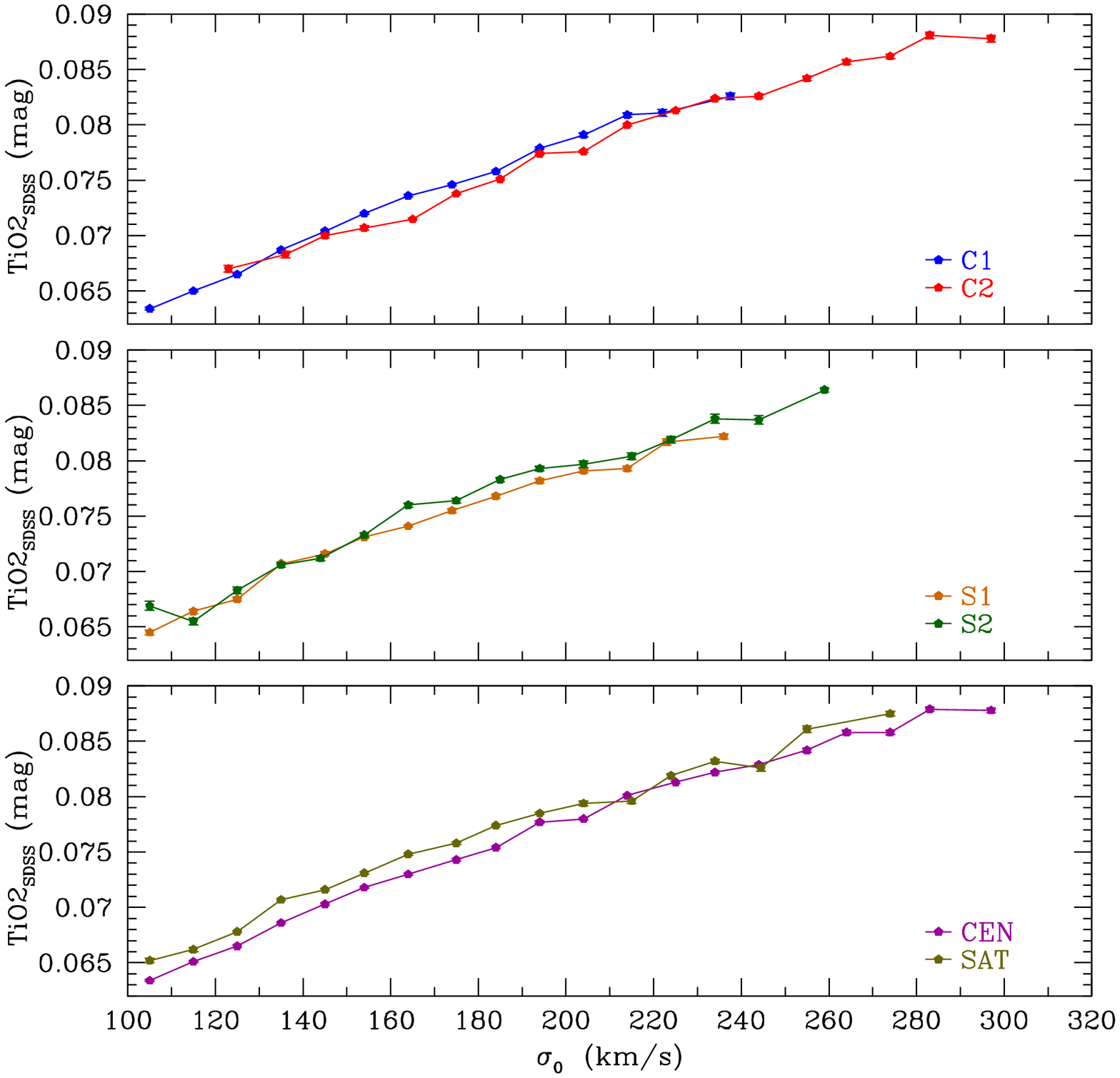}
\caption{Trend of the $\rm TiO2_{SDSS}$ line with $\sigma_0$. $\rm TiO2_{SDSS}$ is mainly used to fit the slope of the IMF and $\rm [Ti/Fe]$, jointly with $\rm TiO1$ and $\rm Fe4531$.}
\label{fig:indTiO2}
\end{figure}

Figs.~\ref{fig:indC}-\ref{fig:indTiO2} show the trends of different line strengths used in the present work as a function of galaxy velocity dispersion, for different subsamples of ETGs. The leading parameters affecting different line indices are summarized in Table~\ref{table:ind}. Differently from the results shown in the present work, where the observed line strengths have been fitted with models smoothed at the nominal $\sigma_0$ of each spectrum, the equivalent widths shown here have been corrected to a reference velocity dispersion of $\sigma=200$~km/s, which lies roughly in the middle of our $\sigma_0$ range. This correction is done to permit a direct comparison between bins of different velocity dispersion and has been performed in the following manner:\par
\begin{itemize}
\item We select five MIUSCAT model spectra, four with a Kroupa-like IMF ($\rm \Gamma_b=1.3$) and with young ($7$ Gyr) and old ($14$ Gyr), as well as solar and super-solar ($\rm [Z/H]=0.22$) metallicities. The fifth model is constructed using a super-solar, old population with a bottom-heavy IMF ($\rm \Gamma_b=2.8$).
\item We choose a velocity dispersion sample $\sigma_i=[100, 320]$~km/s, with a step of $10$~km/s between $\sigma_i$ and $\sigma_{i+1}$, and use the resulting velocity dispersions to broaden the model spectra described in the point above.
\item We then measure the equivalent width of all the indices from our five sets of broadened model spectra.
\item The correction of a given index measured from a spectrum with $\sigma_0=\sigma_i$ is taken as the difference between the equivalent width of the index measured in the model spectrum at $\sigma_i=200$~km/s and the equivalent width of the index measured in the model spectrum at $\sigma_i$: $\Delta EW_i=EW_{200,mod}-EW_{\sigma_i,mod}$. Finally, the correction applied to the measured equivalent widths is the median of the five $\Delta EW_i$ calculated from each set of model spectra.
\end{itemize}
We notice that, as expected, the line strenghts of all metallic lines increase with central velocity dispersion, with the exception of the $\rm Ca$ lines, which remain almost constant as a function of $\sigma_0$. The age sensitive \hbo\ line decreases with $\sigma_0$, consistently with what is shown in Sect.~\ref{sec:res}.\par
Furthermore, small differences are present between different subsamples (in particular C1 and C2) consistent with the results shown in Sect.~\ref{sec:c12}. For instance, \mgb\ ($\rm Fe3$) is lower (higher) for C2 with respect to C1, which is consistent with our result showing that C2 ETGs have slightly lower \mgfe\ and higher metallicity than their C1 counterparts.\par
 Finally we also notice, as described in Sect.~\ref{sec:probC12}, that the index $\rm NaI8190$ (Fig.~\ref{fig:indNaI}) shows a very clear separation between C1 and C2 at low $\sigma_0$, very similar to what we observed in the behaviour of the IMF slope.
%

\bsp	
\label{lastpage}
\end{document}